\documentclass[useAMS,usenatbib]{mn2e}
\usepackage{graphicx}
\usepackage{setspace}
\title[A Spectral Model of WR 136 and NGC 6888]{A Consistent Spectral Model of WR 136 and its Associated Bubble NGC 6888}
\author[J. Reyes-P\'erez, C. Morisset, M. Pe\~na and A. Mesa-Delgado]{J. Reyes-P\'erez$^{1}$\thanks{E-mail:
jperez@astro.unam.mx (JRP)} , C. Morisset$^{1}$, M. Pe\~na$^{1}$ and A. Mesa-Delgado$^{2}$.
\\
$^{1}$Instituto de Astronom\'ia, Universidad Nacional Aut\'onoma de M\'exico\\
$^{2}$Instituto de Astrof\'isica, Facultad de F\'isica, Pontificia Universidad Cat\'olica de Chile\\}

\begin{document}

\date{}

\pagerange{\pageref{firstpage}--\pageref{lastpage}} \pubyear{}

\maketitle

\label{firstpage}

\begin{abstract}

\indent We analyse whether a stellar atmosphere model computed with the code {\small CMFGEN} provides an optimal description of the stellar observations of WR 136 and simultaneously reproduces the nebular observations of NGC 6888, such as the ionization degree, which is modelled with the {\small pyCloudy} code. All the observational material available (far and near UV and optical spectra) were used to constrain such models. We found that even when the stellar luminosity and the mass-loss rate were well constrained, the stellar temperature $T_*$ at $\tau$ = 20, can be in a range between 70 000 and 110 000 K. When using the nebula as an additional restriction we found that the stellar models with $T_* \sim$ 70 000 K represent the best solution for both, the star and the nebula. Results from the photoionization model show that if we consider a chemically homogeneous nebula, the observed N$^{+}$/O$^{+}$ ratios found in different nebular zones can be reproduced, therefore it is not necessary to assume a chemical inhomogeneous nebula. Our work shows the importance of calculating coherent models including stellar and nebular constraints. This allowed us to determine, in a consistent way, all the physical parameters of both the star and its associated nebula. The chemical abundances derived are 12 + log(N/H) = 9.95, 12 + log(C/H) = 7.84 and 12 + log(O/H) = 8.76 for the star and 12 + log(N/H) = 8.40, 12 + log(C/H) = 8.86 and 12 + log(O/H) = 8.20. Thus the star and the nebula are largely N- and C- enriched and  O-depleted.
\end{abstract}


\section{Introduction}

NGC 6888 is an emission nebula associated with the Wolf-Rayet star WR 136 with spectral type WN6(h), and it is a typical example of a ring nebula resulting from the continuous interaction between a strong stellar wind and the interstellar medium. Projected on the sky, this nebula seems to be a filamentary and clumpy gaseous ellipse emitting mainly in H$\alpha$ and optical [N {\small II}] lines, with an axis-size of $12 \times 18$ arcmin \citep{b33}, and it is surrounded by a homogeneous sphere which emits mainly in [O {\small III}] 5007 \AA\ \citep{b34,b19}. The combination of its filamentary morphology and its nitrogen enhanced abundance suggests that the ejections, coming from a previous evolutionary stage of the star WR 136, have contributed in an important way to the formation of the ring nebula that we see today \citep{b5,b35,b18}.

Both the nebula NGC 6888 and its central star WR 136 have been already studied in several wavelength ranges from X-rays to radio emission. For example, the stellar parameters, such as the stellar temperature, mass-loss rate and terminal velocity, have been calculated by \citet{b3} and \citet{b8} by using stellar atmosphere codes and comparing the results with observed ultraviolet, optical and infrared stellar spectra; while the wind velocity law and clumpiness were studied by \citet{b13}. \citet{b36} showed that WR 136 emits in X-ray energies. In particular, these authors found soft emission ($kT \sim 0.56$ keV) coming from outside the wind and hard emission ($kT \sim 2.64$ keV) coming from the gas inside the wind.

The complete hot bubble in the interior of NGC 6888 was observed with the \textit{Space Telescope Suzaku} \citep{b28}, confirming the existence of nitrogen enriched X-ray emitting plasma which consists of two thermal components, one emiting soft X-rays and the other one, hard X-rays. Based on this, the authors proposed a new X-ray emission mechanism based on the thermal interaction between heavy particles, given that the known mechanisms are not capable to explain simultaneously these two X-ray emitting components. However, as these authors explain, it is necessary to have better observational restrictions to confirm this idea. 

On the other hand, complete hydrodynamical simulations performed by \citet{b24}, attempting to reproduce the X-ray observations, indicate that the nebular morphology plays a key role in the presence of X-rays in ring nebulae. These authors conclude that the X-ray emission is higher in bubbles with shell-like morphology than the one present in clumpy filamentary bubbles. However, as the authors indicate, the stellar evolutionary models are not able to reproduce simultaneously the observed stellar parameters and the spectral signatures of the observed X-ray luminosity.

The physical properties in different zones of NGC 6888, such as electronic temperature and density, expansion velocity and chemical composition, have been investigated by several authors since \citet{b37} to recently \citet{b38}. EV92 made a chemodynamical study and they found that NGC 6888 has a chemical composition revealing the material processed in the interior of a star with an initial mass between 25 M$_{\odot}$ and 40 M$_{\odot}$. These results were confirmed by M-D14 who detected for the first time the C {\small II} 4267 \AA\ recombination line, allowing to estimate the C abundance. In this way they presented a complete study of the trace of CNO cycle in the nebula.

FM12 made a very complete study of NGC 6888 through integral field spectroscopy observations performing 1D and 2D analyses. Their results show indeed that NGC 6888 has a complex structure with shells and filaments moving with different velocities. In particular they proposed an ``onion-like" morphology composed mainly by three shells. Concerning the plasma properties, they found that the electron density ranges from less than 100 to 360 cm$^{-3}$ and the electron temperature presents variations from $\sim$7 700 to $\sim$10 200 K. In addition, they reported important variations in the nitrogen abundance across the nebula, the N/O ratio being much higher in the inner region. Recently \citet{b38} derived N/O values from the infrared [N {\small III}] 57-$\mu$m and [O {\small III}]-88 $\mu$m lines and their results would indicate that the edge region of NGC 6888, over the major axis, has lower N abundance with respect to the inner region near the star. 

All the previous works focus only on the modelling of the star or the nebula but not both simultaneously. Further, when a single study of these systems is made (star or nebula) it is difficult to assign a value to the physical parameters only based on generic data. This is why we focus in a semiempirical method based in the self-consistent modelling of the star and its associated nebula.

It is logical to proceed in this way because the nebula associated to the star is naturally an additional restriction to the stellar atmosphere model, because the nebula can tell us about the ionizing ultraviolet photons coming from the star and then reprocessed and reemitted with low energy through recombination and collisional processes. In this way we can realize how well a stellar model can produce the ionization stage of the nebula and for this purpose we build a photoionization model of NGC 6888 based on spectroscopic observations where the modelled stellar spectral energy distribution (SED) is used as the ionization source.

This methodology will give us the advantage to obtain more constrained and coherent physical values for all the components of the entire system than those resulting from a single study of each component as it was made until now. The results from this work will help to better constraint other kind of studies as models of stellar evolution, astrophysical bubble formation and, in particular, the presence of X-rays in the bubbles since this hot gaseous component imposes an additional and important restriction to understand the whole system, given that is there where the stellar wind energy is stored.
 
This paper is structured as follows: In Section 2, observations and data reduction process are described. Sect. 3 contains the information of the line flux measurements and the error estimates. In Sect. 4 we describe the stellar and nebular models as well as our methodology to constraint the free parameters. A discussion about our best self-consistent model is presented in Sect. 5. Finally, in Sect. 6, we present our conclusions.

\begin{figure}
\includegraphics[width=0.5\textwidth]{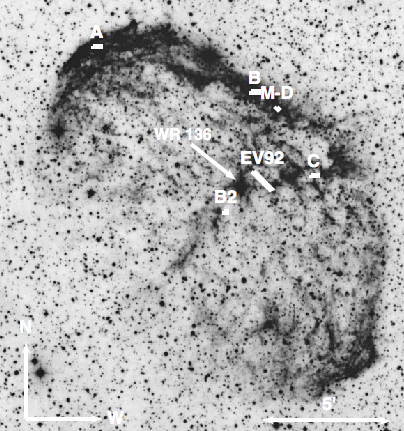}
 \caption{Slits positions where the observations were performed on NGC 6888 including the positions of M-D14 labeled as M-D, B2 by FM12 and the position of EV92. Red filter image from the STScI-DSS I/II server.}
 \label{fig:slit}
\end{figure}

\begin{table*}
\centering
\caption{Log of observations}
\label{tab:logob}
\begin{tabular}{lccccc}
\hline
\hline
Observation ID & A. R  & Dec.  & Date & Exp. Time & Slit dimensions\\
                         &J2000 & J2000 &         &  (s)           & (arcsec)\\
\hline
NGC 6888A &20:12:36.5& 38:26:56.5 & 1st sept. 2011 & 4$\times$1800 & 4$\times$13.3\\
NGC 6888B&20:12:03.7 &38:25:07.8 &  30 Aug. 2011& 4$\times$1800& 4$\times$13.3\\
NGC 6888C&20:11:51.2 &38:21:45.8&   30 Aug. 2011&4$\times$1800& 4$\times$13.3\\
WR 136& 20:12:06.5 &38:21:17.7 & 28 Aug. 1997& 3$\times$300& 9.3$\times$13.3\\
HR 7596 & 19:54:44.7 &00:06:25.1 & 1st sept. 2011 & 3$\times$420& 9.3$\times$13.3\\
\hline
\hline
\end{tabular}
\end{table*}

\section{Observations and data reduction}

The main analysis in this work is based on the optical observations carried out the night 1997 August 28, for the star WR 136; and the 2011 August 31 and September 1, for the nebula and the standard star HR 7596. We used the REOSC echelle spectrograph \citep{b16} attached to the 2.1-m telescope at the Observatorio Astron\'omico Nacional (OAN) in San Pedro M\'artir, B. C. M\'exico. With this spectrograph we obtained high-resolution spectra ($R\sim 20 000$) with a wavelength coverage from 3700 \AA\ to 7300 \AA.

Three different positions of the nebula were observed using a slit with size 4 $\times$ 13.3 arcsec oriented east-west. For each position we obtained four spectra with an exposure time of 1800 s each. The slit positions are shown in Figure~\ref{fig:slit}. Together with slit positions of other authors that will be described in the following sections.

To perform the spectroscopy of WR 136 and the standard star the slit size was 9.3 $\times$ 13.3 arcsec. For each star three spectra were obtained, with a exposure time of 420 s in the case of the WR star and 300 s for the standard star. The log of observations is shown in Table~\ref{tab:logob}.
 
After the bias correction was made, the corresponding different exposures were combined and then background and cosmic rays were subtracted using the \textit{filter/cosmic} task of the astronomical software {\small MIDAS}\footnote{http://www.eso.org/sci/data-processing/software/esomidas}. Then, once the wavelength calibration was performed (using a ThAr lamp), the orders of the echelle images were extracted.

To correct the extracted orders by the blaze function and to make the flux calibration in the case of the nebular spectra, we modelled the star HR 7596. First the individual orders were normalized to a curve which is theoretically the convolution between the blaze function and the sensitivity function. In this way, the normalized orders were merged to obtain a complete normalized stellar spectrum to be modelled with the code {\small SPECTRUM} \citep{b7}. This code assumes a LTE atmosphere with a plane-parallel geometry and is adapted for modelling standard stars with a spectral type between B and M. The spectrum can be calculated in a normalized way or the model can give the emergent flux per wavelength unit. First we obtained the normalized modelled spectrum to directly compare it with the observed spectrum and once this one was reproduced, the not normalized spectrum was calculated.

The model spectrum of the standard star was then reddened considering the interstellar and atmospheric extinction as well as the radiation dilution. In this way, we constructed the synthetic standard star spectrum. Then, the extracted orders of the standard star observation can be divided by this spectrum to obtain the single order function where the information about the instrumental sensitivity and flux calibration is contained. To correct for the CCD sensitivity and to flux-calibrate the nebular spectra, these orders were divided by their corresponding single order function. Finally the corrected extracted orders were analysed as individual spectra. These resulted spectra have a total exposure time of two hours.

In the case of the orders from the WR 136 frames, we proceeded with the rest of the reduction similarly as we made with the standard star and given that observations of the standard star and WR 136 were not made at the same date, the resulting optical spectrum was just normalized to the continuum.

In addition to the optical spectra of WR 136, we analyzed a spectrum in the UV region between 900 \AA\ and 1200 \AA\ obtained from the \textit{FUSE} satellite observational program C097 where some interesting and useful lines, such as S V $\lambda$1063 and P V $\lambda$1118 lie. To extend the wavelength range to 3300 \AA\ we retrieved several reduced IUE spectra\footnote{see https://archive.stsci.edu/iue/file\_formats.html for more information}. To improve the signal to noise ratio we combined the individual IUE short-wavelength high-dispersion spectra SWP57618, SWP57622, SWP57625, SWP57626, SWP57630, SWP57635, SWP57638, SWP57645, SWP57647 and SWP57663; and the individual IUE long wavelength high dispersion spectra LWP32398, LWP32407, LWP32409, LWP32413, LWP32416, LWP32418, LWP32423, LWP32440, LWP32441 and LWP32446.

\begin{figure*}
\includegraphics[width=1.0\textwidth]{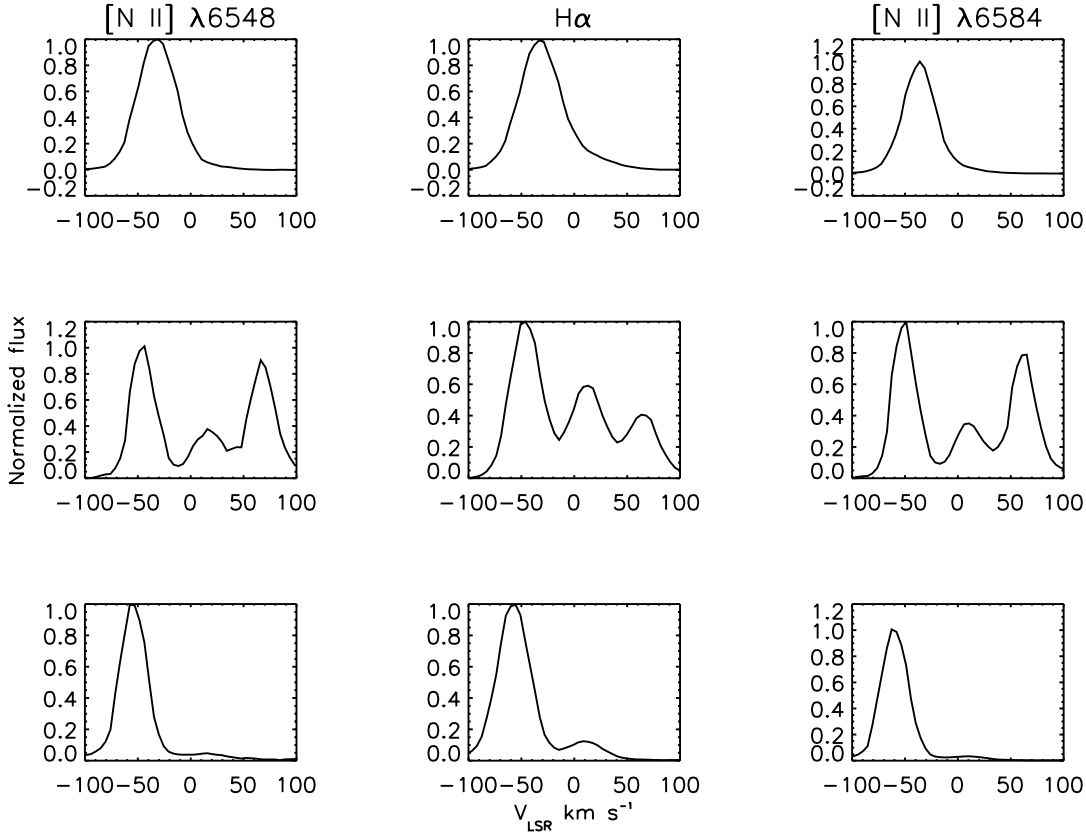}
\caption{The H$\alpha$ and [N {\tiny II}] 6548, 6583 lines are shown in radial velocity space for positions A (top), B (middle) and C (bottom). The velocities are measured respect to the Local Standard of Rest. Spectra at position A only present one component at -36 km s$^{-1}$, while the spectra at position B show two components corresponding to the expanding shell (bluest and reddest components) at -45 and 65 km s$^{-1}$ and a central component corresponding to the H {\tiny II} region around NGC 6888 at 15 km s$^{-1}$. Finally in the spectra at position C two components are detected at -57 and 12 km s$^{-1}$.}
\label{fig:specneb}
\end{figure*}

\section{Flux measurement and error estimates}

\begin{table*}
\caption{Fluxes (normalized to F$_{H {\beta}}$ = 100) of the identified lines in the observed spectra, for positions A, B (blue, central and red) and C(blue and central). The flux F$_{H {\beta}}$ is in units of 10$^{-12}$erg cm$^{-2}$ s$^{-1}$.}
\label{tab:fluxes}
\begin{tabular}{llcccccc}
\hline
\hline
$\lambda$ (\AA)& Ion & A & B$_{b}$ & B$_{c}$ & B$_{r}$ & C$_{b}$ & C$_{c}$\\
3726.03 &[O {\small II}]&52.6$\pm$14.3&	...	&	...	&	...	&	...	&	...\\							3728.81 &[O {\small II}]& 62.6$\pm$9.5&	...	&	...	&	...	&	...	&	...\\							
3970.07 &H$_{\epsilon}$ & 16.8$\pm$6.1&	...	&	...	&	...	&	...	&	...\\						
4101.73 &H$_{\delta}$& 28.1$\pm$13.0& 19.1$\pm$11.0&	...	&	...	&	24.2$\pm$9.4&	...\\					
4340.46 &H$_{\gamma}$& 47.1$\pm$5.7& 48.8$\pm$13.4& 26$\pm$20.5& 36.9$\pm$21.4& 47.7$\pm$ 3.5&	43$\pm$24.0\\				
4471.47 &He {\small I}& 6.0$\pm$1.2& 4.2$\pm$1.7& 3.5$\pm$2.4&2.9$\pm$3.3	& 7.5$\pm$2.1&	...\\						
4958.91 &[O {\small III}]& 23.8$\pm$2.4& 97.0$\pm$6.2& 72.9$\pm$24.1& 101.2$\pm$22.8 & 52.1$\pm$4.5&	77.8$\pm$47.5\\			
5006.84 &[O {\small III}]& 67.9$\pm$5.6& 289.2$\pm$16.3& 210.9$\pm$66.4& 250.2$\pm$57.2& 154.2$\pm$9.2& 230.2$\pm$101.5\\			
5015.67 &He {\small I}& 4.7$\pm$1.8& 6.0$\pm$1.8&	...& 44.7$\pm$27.2& 4.7$\pm$3.1&	6.8$\pm$5.3\\							
5754.64 &[N {\small II}]& 3.8$\pm$1.2& 2.3$\pm$1.3&	...	&	...	& 2.3$\pm$0.9&	...\\							
5875.61 &He {\small I}& 19.9$\pm$2.3& 21.3$\pm$2.5& 14.5$\pm$6.7&	...	& 23.5$\pm$2.1& 18$\pm$11.9\\						
6312.10&[S {\small III}]& ...&	2.2$\pm$1.3&	...	&	...	& 1.4$\pm$0.6&	...\\								6548.04 &[N {\small II}]& 168.4$\pm$11.5& 37.1$\pm$2.8& 28.1$\pm$10.4& 81.4$\pm$18.8& 70.9$\pm$3.5& 40.7$\pm$17.5\\
6562.80 &H$_{\alpha}$& 285.9$\pm$18.3& 286.4$\pm$14.1& 285.9$\pm$84.4& 286.2$\pm$58.8& 286$\pm$10.8&286$\pm$92.14\\			
6583.46 &[N {\small II}]& 502.1$\pm$31.9&	106.8$\pm$6.1& 75.5$\pm$24.4& 236.6$\pm$49.7& 227.0$\pm$9.7& 123.6$\pm$46.9\\			6678.15 &He {\small I}& 6.4$\pm$1.2& 6.4$\pm$2.8& 3.2$\pm$2.6& 8$\pm$5.7& 5.2$\pm$0.7&	...\\								
6716.44 &[S {\small II}]& 22.9$\pm$2.5& 7.0$\pm$2.1& 24.9$\pm$10.0& 8.4$\pm$5.5& 7.4$\pm$0.8& 30.1$\pm$13.1\\					
6730.81 &[S {\small II}]&19.9$\pm$1.9& 5.7$\pm$2.4& 17.8$\pm$7.1& 7.1$\pm$4.5& 6.3$\pm$1.2& 19.9$\pm$9.6\\						7065.17 &He {\small I}& 3.2$\pm$1.0& 6.0$\pm$2.6&	...	&	...	& 6.4$\pm$0.8&	...\\							
7135.80&[Ar {\small III}]& 9.2$\pm$1.1& 15.6$\pm$2.7& 9.0$\pm$5.8& 12.4$\pm$6.7& 10.0$\pm$1.1&	...\\							
\hline
\hline
& F$_{H {\beta}}$  & 4.70$\pm$0.05 &0.110$\pm$0.004 & 0.077$\pm$0.004 & 0.037$\pm$0.005 & 0.565$\pm$0.023 & 0.076$\pm$0.022\\
& c$_{H {\beta}}$ & 0.70$\pm$0.02& 0.38$\pm$0.07& 0.36$\pm$0.11& 0.77$\pm0.17$ & 0.0 & 0.03$\pm$0.45\\ 
\hline
\hline
\end{tabular}
\end{table*}

In Figure~\ref{fig:specneb} we show the resulting H$\alpha$ and [N {\small II}] 6548, 6584 \AA\ lines present in each spectrum of NGC 6888. In this figure it is apparent that the spectra show line emissions coming from three components at different velocities. This is more obvious for the zone B in whose spectrum we can clearly distinguish these components located at around $-$45, 15 and 65 km s$^{-1}$. According to the wavelength calibration performed, the radial velocities have an accuracy of $\pm$12 km s$^{-1}$.

In the spectrum of position C only two components are present with radial velocity of $-$57 and 12 km s$^{-1}$. Probably the redshifted one is also present but very weak. In the case of position A, corresponding to the nebular edge, it is only evident the more intense blueshifted component at around $-$36 km s$^{-1}$. We can see that the line profiles are not gaussian at all, they exhibit a wing on the red side where surely the others components, not resolved, are contained.

The radial velocities of the blueshifted and redshifted components agree with the values in the literature. For example, in a position near the star, EV92 detected two components at $-$64 and 78 km s$^{-1}$ and M-D14 reported two blueshifted components at $-$60 and $-$25 km s$^{-1}$, all of them associated with the emission of NGC 6888. Recently, performing a 2D study in a region close to our position A, FM12 found that the emission of the [N {\small II}] 6584 \AA\ line comes from plasma with radial velocities between $-$40 to 30 km s$^{-1}$, in agreement with our value at position A. 

The kinematical component at $\sim$15 km s$^{-1}$ is surely associated with the extended H {\small II} region around NGC 6888 and it has been also identified by EV92 and M-D14 with a velocity of 18 km s$^{-1}$ and 12 km s$^{-1}$, respectively.

In general, we found radial velocities which follow the trend previously reported, i.e, the highest values appear near the star and they decrease as moving away from the centre of the nebula.

For the spectra at positions B and C we measured the line fluxes by fitting a gaussian profile to each kinematical component, but for the spectrum at position A, all the emission was integrated for each identified line. The errors assigned to our measurements correspond to a two sigma value over the mean noise measured at each side of the lines detected.

The fluxes were then dereddened with a logarithmic extinction coefficient c$_{H\beta}$ determined from the H$\alpha$/ H$\beta$ ratio using \citet{b1} extinction law. The uncertainties in c$_{H\beta}$ are the result of the flux-error propagation.

The c$_{H\beta}$ value at position A (0.70) agrees with the mean value of 0.76 obtained by FM12 in a region close to this observation. In the case of position B the extinction values are 0.36, 0.38 and 0.77 for the blueshifted, foreground and redshifted components, respectively. The first and second values are similar to the value 0.34 obtained by M-D14 for a region near our position B.

The high c$_{H_{\beta}}$ value of the redshifted component seems to indicate that the emission coming from the back shell of NGC 6888 is more reddened surely by the presence of inner extinction. This is supported by the extinction value for WR 136, $A_{V} = 1.73$ mag, reported by \citet{b39}. Using $R_{V} = 3.1$ and $E_{B-V} = 0.692\times c_{H\beta}$, as derived from the extinction law mentioned above, we find c$_{H\beta} = 0.80$ for the star. Given that WR 136 is embedded in NGC 6888, inner extinction could be contributing to redden its radiation. 

Finally, observations of position C do not seem to be reddened. The only observations to compare with a neighbour zone are those performed by EV92. They found $c_{H\beta} = 0.28$, which is the lowest value reported in the works cited above, indicating this region of NGC 6888 does not present high extinction. 

In Table~\ref{tab:fluxes} the whole set of identified lines with their respective dereddened flux-integrated values are shown. In this table we present one set of values for position A, three (B$_{b}$, B$_{c}$ and B$_{r}$) for position B, and two (C$_{b}$ and C$_{c}$) for position C, where the subscripts ``b'', ``c" and ``r" correspond to the blueshifted, central and redshifted components, respectively.

\begin{table*}
\centering
\begin{minipage}{140mm}
\caption{Stellar parameters}
\label{tab:w136}
\begin{tabular}{lcccccccccc}
\hline
\hline
Model & $R_{\ast}$\footnote{{\small $R_{\ast}$ and $T_{\ast}$ are defined at $\tau=20$}} & log $L$ & log $\dot{M}$ & $V_{\infty}$& $T_{\ast}$ & $T_{2/3}$ & $\beta$ & $f_V$& X & Y\\
        & [$R_{\odot}$] & [$L_{\odot}$] & [$M_{\odot}$yr$^{-1}$] & [km s$^{-1}$] & [kK] & [kK] & & & &\\
\hline
This work \textit{star1} & 1.50 & 5.45 & -4.63 & 1550 & 110 & 55 & 2 & 0.1 & 0.016&0.95\\
This work \textit{star2}& 2.10 & 5.40 & -4.95 & 1550 & 90 & 48.30& 1 & 0.175 & 0.068 & 0.91 \\
This work \textit{star3}& 3.38 & 5.40 & -4.95 & 1550 & 70 & 45.91& 1 & 0.25 & 0.12 & 0.86 \\
This work \textit{star4}& 3.38 & 5.40 & -4.95 & 1550 & 70 & 42.26& 2 & 0.25 & 0.12 & 0.86 \\
Hamann et al.(2006) & 3.34 & 5.40 & -4.95 & 1600 & 70.8 & ...& 1 & 0.25 & 0.12 & 0.86\\
Crowther \& Smith (1996) & 5.00 & 5.30 & -3.91 & 1750 & 55.7 & 28\\
\hline
\hline
\end{tabular}
\end{minipage}
\end{table*}

\begin{figure*}
\includegraphics[width=1.0\textwidth]{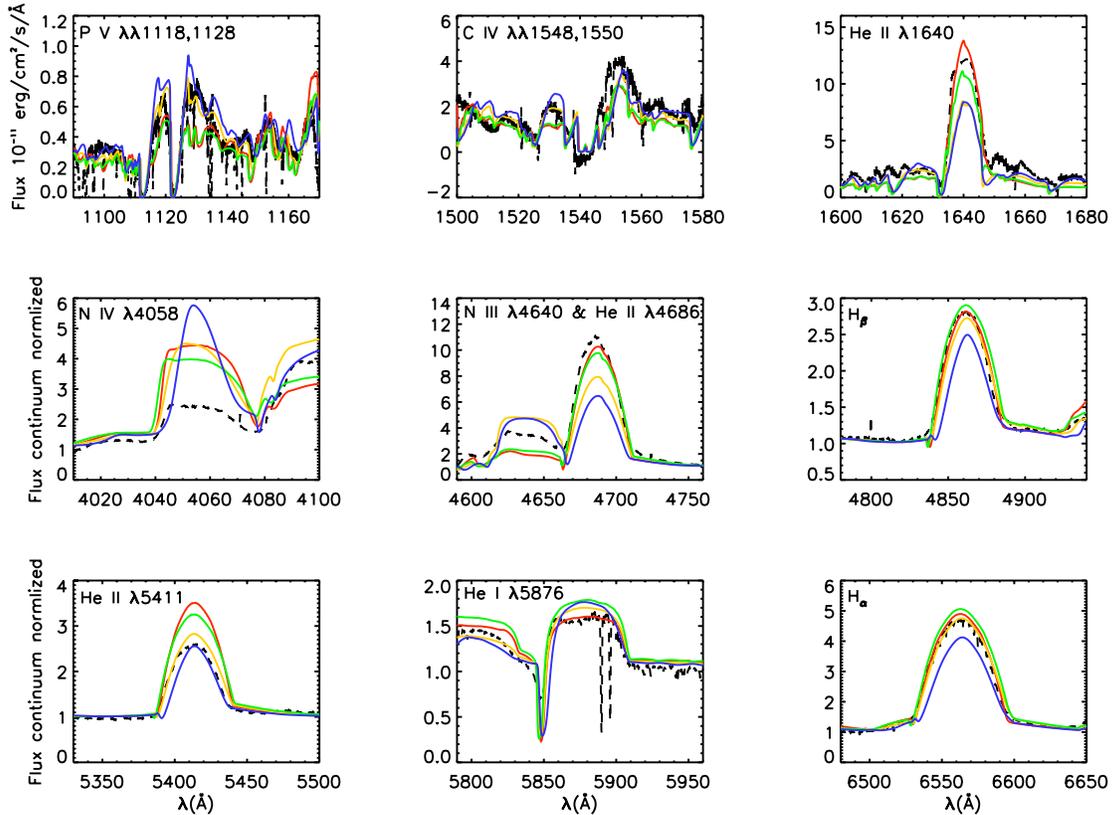}
 \caption{Selected lines from the observed spectra to constrain the models. In each plot the dashed black line is the observed spectrum; red, green, yellow and blue lines are respectively the \textit{star1, star2, star3} and \textit{star4} models obtained through the {\small CMFGEN} code but reddened. The optical spectrum is shown as flux-normalized.}
 \label{fig:steli}
\end{figure*}

\begin{figure*}
\includegraphics[width=1.0\textwidth]{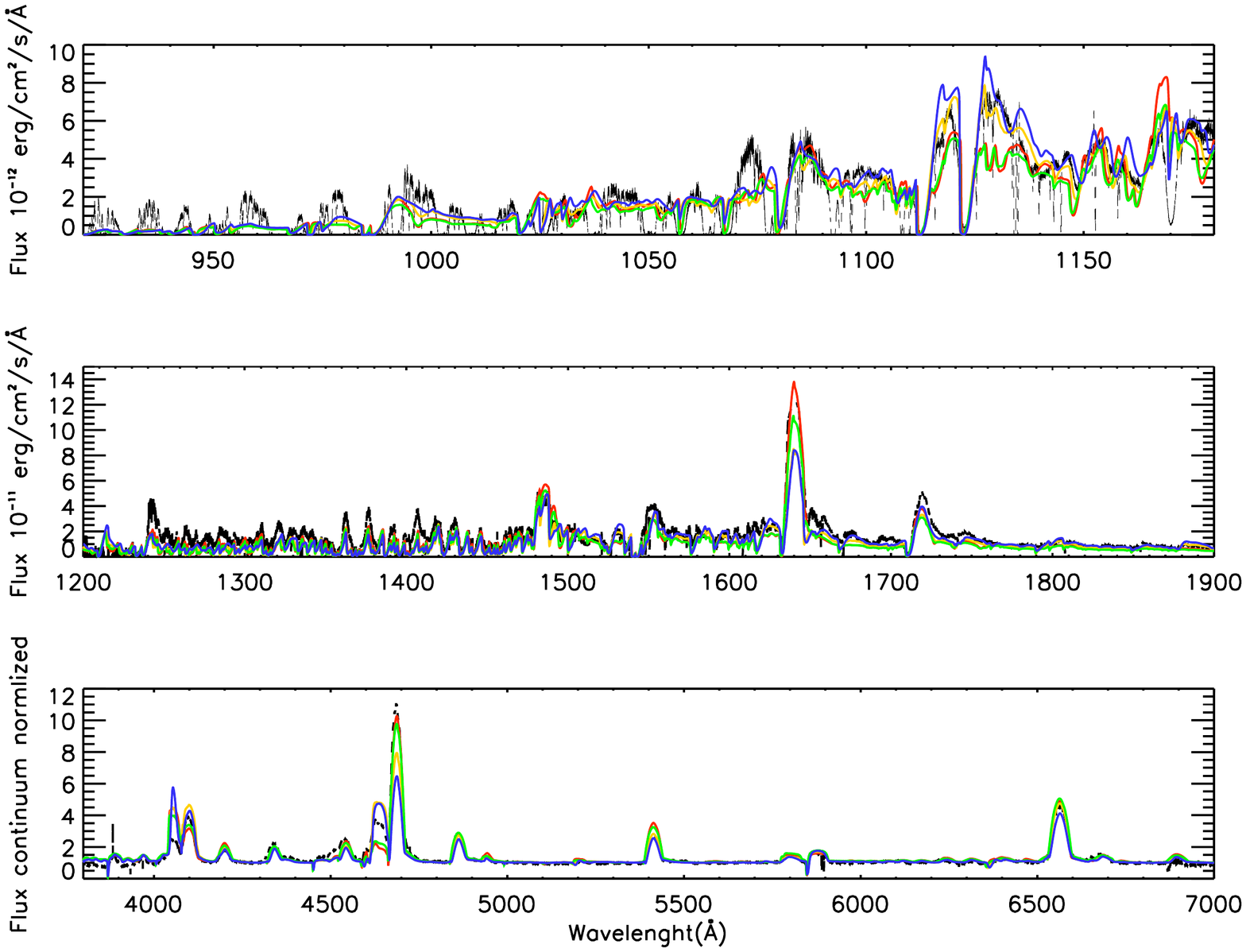}
 \caption{General view of the \textit{star1, star2, star3} and \textit{star4} models compared with the observed spectra (dashed black line). From up to down: the \textit{FUSE} FUV spectrum, the \textit{IUE} UV spectrum and the continuum-normalized optical spectrum. The colors are as in Fig.~\ref{fig:steli}}
 \label{fig:stelspec}
\end{figure*}

\section{The models}
\subsection{The stellar model}

To model the observed stellar spectra of WR 136, in the UV and  optical ranges, the {\small CMFGEN} code \citep{b10} was used. {\small CMFGEN}  (CoMoving Frame GENeral) is a non-LTE  line-blanketed atmospheric code designed for spectral analysis of stars with stellar winds. It solves the radiative transfer and statistical equilibrium equations considering spherical symmetry. This code is adequate to model the atmosphere of massive stars with an extended wind that is, in general, larger than the hydrostatic radius of the star.
 
To start-up {\small CMFGEN} code it is required to set the stellar hydrostatic structure through input parameters such as the gravity and stellar radius or, in an equivalent way, to set a velocity law that approaches in depth to the hydrostatic structure. In this study we chose the later option and used the velocity law given by:\\

\begin{equation}
V(r) = \frac{V_{o}+(V_{\infty}-V_{o})(1-R_{\ast}/r)^\beta}{1+(V_{o}/V_{core} - 1)exp([R_{\ast}-r]/h_{eff})},
\end{equation}

\noindent as proposed by \citet{b11}. In expression (1) $V_{core}$ is the velocity at the hydrostatic stellar radius $R_{\ast}$, $V_{o}$ is the photospheric velocity which controls the match between the hydrostatic atmosphere and the wind, $V_{\infty}$ is the terminal velocity, $h_{eff}$ is the scale height of photosphere, in units of $R_{\ast}$, which specifies the density structure at low velocities and $\beta$ is the so-called velocity law exponent. The values used for all our models for $V_{core}$, $V_{o}$ and $h_{eff}$ were 1 km s$^{-1}$, 100 km s$^{-1}$ and 0.002, respectively, as it is suggested for WR stars\footnote{See the CMFGEN Documentation and WR models at kookaburra.phyast.pitt.edu/hillier/web/CMFGEN.htm}. The $\beta$ values used for the models were 1 or 2, which are usual values for this kind of stars \citep{b13}. 

Then, once the code calculates the hydrostatic values, it solves the radiative transfer equations in the wind zone and the result is the emergent flux (as emitted at 1 kpc from the observer) as a function of the wavelength. To fit the UV spectra, we used for the modelled spectrum a distance of 1.45 kpc, based on the \textit{HIPPARCOS} parallax \citep{b26}, and a final extinction value $E_{B-V}$ of 0.52 (or c$_{H\beta}$ = 0.75) which agrees with the value by \citet{b39} shown in \S 3. In this way we fit the stellar continuum and the absorption feature around 2200 \AA.  For the optical spectrum, the model was analysed using the continuum normalized spectrum.

For the stellar models computed in this study, the input values used to constrain the model atmosphere for  WR 136 were the mass-loss rate $\dot{M}$, the terminal wind velocity $V_{\infty}$, the stellar luminosity $L$, the hydrostatic radius of the star $R_{\ast}$ (at optical depth $\tau$ equal to 20) and the chemical composition.

According to the stellar wind theory, it is expected that the recombination line luminosity $L_{\lambda}$ is proportional to the mass-loss rate through the relation $\dot{M}\sim V_{\infty}\sqrt{L_{\lambda}}$, therefore the H$\alpha$ recombination line was used to adjust the value of $\dot{M}$ \citep{b15}.
 
On the other hand, when we consider an extended stellar atmosphere, it is not possible to assign a surface temperature to the star as the temperature varies with the radius, therefore the temperature cannot be a global input value in the model. However, it is possible to define a temperature value $T_{\ast}$ for  $\tau$ = 20 and such a parameter is set in the model through the luminosity and the stellar radius with the common expression $L = 4 \pi \sigma R_{\ast}^{2}T_{\ast}^{4}$, where $\sigma$ is the Stefan-Boltzmann constant. Therefore, for constructing the model, the $L$ and $R_{\ast}$ values were changed to fit the line intensities corresponding to consecutive ionization degrees of the same element, such as the He {\small I} and He {\small II} lines.

While constructing a model and when the He lines temperature indicators (He {\small II} $\lambda$4686 and He {\small I} $\lambda$5876) pointed that we were very close to the best fit, we found that the predicted  N {\small III} $\lambda$4640 line was weaker than the observed line, while the predicted N {\small IV} $\lambda$4058 was more intense than the observation, implying the temperature should be smaller (see Fig.~\ref{fig:steli}). It has been stated in the literature that the ion N {\small III} is very dependent on the detail of its atom model \citep{b11}. We tried to correct this issue by using a more complex N {\small III} ion model, i.e., more levels were considered when decollapsing a superlevel. However, the problem was not solved although the fit of these nitrogen lines improved slightly.

The value of the terminal velocity was constrained by fitting the blue edge of the saturated C {\small IV} $\lambda$1550 P-Cygni profile. The obtained value (1550 km s$^{-1}$) fits well other P-Cygni profiles such as S {\small IV} $\lambda$$\lambda$1063,1073 and P {\small V} $\lambda$$\lambda$1118,1128 in the far UV spectrum of the star (see Fig.~\ref{fig:steli}) and it is in excellent agreement with values reported in previous works. \citet{b21} used the common diagnostic line, the saturated UV C {\small IV} resonance doublet at 1548, 1550  \AA\, and obtained $V_\infty = 1605$ km s$^{-1}$; \citet{b4} derived $V_\infty = 1660$ km s$^{-1}$ by using the He {\small I} 10830 \AA\ line width; HGL06 measured $V_\infty = 1600$ km s$^{-1}$ by fitting their model to their observed spectrum; \citet{b3} determined a $V_\infty = 1750$ km s$^{-1}$; finally, through an IR study, \citet{b12} used the Ca {\small IV} 3.207 $\mu$m forbidden line width, reporting $V_\infty = 1490$ km s$^{-1}$. As our fit was done measuring the velocity at the edge of the satured component of the blue wing, instead of measuring it where the wing reaches the continuum, this renders a terminal velocity $\sim 200$ km s$^{-1}$ lower than when measured at the continuum. This excess in velocity can be explained invoking a turbulent velocity or by the clumping factor $f_{V}$ of the wind.

The chemical composition in the models was initially set at typical WN composition \citep{b25} and it was modified to generate some of the models shown here. The final H and He abundances (X, Y by mass-fraction) employed for all the stellar models are shown in Table~\ref{tab:w136}, while for the other elements such as C, N, O, S and Fe, the abundances derived from {\small CMFGEN} are shown in Table~\ref{tab:abun} to compare then with the resulting nebular abundances from the modelled nebula.
 
As the modelled spectral features do not respond linearly to the input parameters, once a value is found it is not necessarly fixed but only constrained while changing the other input values. Therefore we calculated a grid of models in order to find the more appropriated models which reproduce most of the observed spectral signatures. 

To properly fit the luminosity through the stellar continuum without changing the stellar temperature, it was necessary to use the transformed radius

\begin{equation}
R_t = R_\ast\left(\frac{V_\infty/2500}{\dot{M}\sqrt{D}/10^{-4}}\right)^{2/3},
\end{equation}

\noindent (where $D = f_{V}^{-1}$) because it has been shown that two models with the same $R_{t}$ and $T_{\ast}$ values have the same spectral behaviour \citep{b23}.

From the grid of models calculated by varying the input parameters we obtained \textit{star1} which fits very well most of the stellar spectral characteristics. This model is presented in Table~\ref{tab:w136}. In addition we also show in this table models from the literature, as HGL06 and \citet{b3} ones. The mass-loss rate value of \textit{star1} is slightly higher but of the same order as  HGL06 value and it is lower than the value published by \citet{b3}. We are confident in the mass-loss rate value of our models because as it can be seen in Figure~\ref{fig:steli} the recombination lines, like the Balmer's series in the optical spectrum, are very well fitted.

In Figure~\ref{fig:steli} we can see that the He {\small II} $\lambda$1640 and $\lambda$5411 lines are overestimated in \textit{star1} only by a factor $\sim1.3$. It is important to say that it has been stated in the literature that the He lines in an observed spectrum cannot be fitted simultaneously \citep{b11}. From Figure~\ref{fig:stelspec} we can also see that \textit{star1} reproduces in general very well the observed spectra from the UV to the optical wavelength range.

The main difference between \textit{star1} and the model reported by HGL06 is the stellar radius $R_{\ast}$ which in our case has a low value of 1.5 $R_{\odot}$ (compared to the value of 3.34 $R_{\odot}$ by HGL06) leading to a higher stellar temperature $T_{\ast}$ of 110 kK. Additionally the chemical composition, represented by the values of X and Y, is different. These differences have important implications when modelling the nebula around the star (see \S 4.2.2) and are indicating that it could exist more than one solution for a stellar model of a WR star as both, \textit{star1} and HGL06 model, reproduce similarly well the stellar spectrum (see Fig. 68 by HGL06). This apparent degeneracy has already been remarked in a study of WR stars modelling by \citet{b9}. In this work the origin of this degeneracy is attributed to our misunderstanding of the dynamics in the inner wind regions. 

To investigate this degeneracy we searched for other {\small CMFGEN} models that could fit the stellar observations. The results are plotted in Figure~\ref{fig:stelspec} where the general behaviour of the models is compared with the observed spectra.
 
In a simple manner we used interpolated values between the model \textit{star1} and the HGL06 model for the radius, the clumping factor and the hydrogen and helium fractions by mass, X and Y. The result is model \textit{star2} which has an effective temperature of 90 000 K (its characteristics are presented too in Table~\ref{tab:w136}). Model \textit{star2} reproduces the stellar spectrum similarly well as \textit{star1}. It is presented in Fig.~\ref{fig:stelspec}, in green.

We computed two additional models. The first is \textit{star3} which basically uses HGL06 model parameters but it was calculated with the {\small CMFGEN} code. Model \textit{star3} reproduces very well the results of HGL06 model, which shows that {\small CMFGEN} and the Potsdam WR model atmosphere code are equivalent. The last one is model \textit{star4} which is almost the same as \textit{star3} but calculated with the $\beta$ value equal to 2, leading to a lower $T_{2/3}$ value because the wind region becomes more extended. Models \textit{star3} and \textit{star4} reproduce fairly well the observed stellar continuum and many spectral features (lines yellow and blue in Fig.~\ref{fig:stelspec}) although the emission lines indicative of the stellar temperature are worse fitted than with \textit{star1}(see Fig.~\ref{fig:steli}). All the stellar model characteristics are presented in Table~\ref{tab:w136}.

From this, it is then clear that there is not a unique stellar model reproducing the stellar spectrum. As said above, we are in presence of a degeneracy occurring when modelling WR stellar atmospheres and evidently more constraints are required to decide for the best model. Additional constraints can be found by analyzing the ionized nebula.

\begin{table*}
\begin{center}
\begin{minipage}{150mm}
\caption{Physical parameters and ionic and total abundances}
\label{tab:par}
\begin{spacing}{2.0}
\begin{tabular}{lcccccc}
\hline
\hline
Parameter & Position A & Position B$_{b}$ & Position C$_{b}$& EV92 Position & Position B2 (FM12)& M-D14\\
\hline
T$_{e}$[O III](K)& ... & ... & ... & 9500:& (6855$\pm$639)\footnote{\small Temperature derived from \citet{b57} formula.} & 9050$\pm$810\\
T$_{e}$[N II](K)&7800$^{+760}_{-870}$& 11700$\pm$3000 &9100$\pm1150$& 8400$\pm$100 & 8537$\pm$527& ...\\
n$_{e}$[S II] (cm$^{-3}$)& 240$^{+280}_{-120}$ & 180$^{+3750}_{-100}$ & 215$^{+380}_{-130}$& 371 & 108$\pm26$& 480:\\
n$_{e}$[O II] (cm$^{-3}$)& 190$^{+670}_{-110}$ & ...& ...&301&...& 310$\pm$30\\
12+log O$^{+}$/H$^{+}$& 8.12$^{+0.37}_{-0.20}$& ... & ...& 7.54$\pm0.34$ & 7.96$\pm$0.14& 7.45$\pm$0.14\\
12+log O$^{++}$/H$^{+}$& 7.79$^{+0.28}_{-0.16}$& 7.80$^{+0.5}_{-0.26}$ & 7.89$^{+0.26}_{-0.18}$& 7.98$\pm$0.25 & 8.57$\pm0.19$& 8.02$\pm$0.11\\
12+log N$^{+}$/H$^{+}$& 8.31$^{+0.21}_{-0.12}$ & 7.18$^{+0.53}_{-0.21}$ & 7.78$^{+0.20}_{-0.14}$& 7.83$\pm0.30$ & 7.63$\pm0.08$& 7.54$\pm$0.08\\
12+log S$^{+}$/H$^{+}$& 6.34$^{+0.21}_{-0.11}$ & 5.36$^{+0.51}_{-0.19}$ & 5.66$^{+0.21}_{-0.13}$& 5.71$\pm0.23$ & 5.80$\pm0.08$& 5.43$\pm$0.09\\
12+log S$^{++}$/H$^{+}$&...& 6.45$^{+0.50}_{-0.50}$ & 6.75$^{+0.35}_{-0.40}$ & ... & ...\\
12+log Ar$^{++}$/H$^{+}$ & 6.20$^{+0.20}_{-0.12}$ & 6.00$^{+0.48}_{-0.20}$ & 6.12$^{+0.18}_{-0.13}$& ...&...& 6.36$\pm$0.10\\
12+log He$^{+}$/H$^{+}$&11.20$\pm$0.18&11.29$\pm$0.14&11.21$\pm$0.36& 11.24$\pm$0.06 &11.20$\pm$0.06&11.21$\pm$0.03\\
12+log O/H& 8.29$^{+0.34}_{-0.19}$& ... & ...& 8.11$\pm$0.28 &8.66$\pm$0.16&8.20$\pm$0.09\\
12+log N/H&8.48$^{+0.12}_{-0.17}$& ... & ...& 8.40$\pm0.35$ &8.34$\pm0.23$& 8.54$\pm$0.20\\
12+log S/H&7.15$^{+0.20}_{-0.12}$& ... & ...& ... & ...& 6.77$\pm$0.20\\
12+log Ar/H&6.47$^{+0.20}_{-0.12}$& ... & ...&...&...& 6.41$\pm0.11$\\
log N/O &0.19$^{+0.58}_{-0.32}$ & ... & ... & 0.29$\pm$0.30& -0.33$\pm$0.28& 0.22$\pm$0.22\\
\hline
\hline
\end{tabular}
\end{spacing}
\end{minipage}
\end{center}
\end{table*}

\begin{table*}
\begin{center}
\caption{Atomic data}
\label{tab:atom}
\begin{tabular}{lll}
\hline
\hline
       & Transition & Collisional\\
Ion & probabilities & strengths\\
\hline
N$^{+}$ & \citet{b44} & \citet{b53}\\
             & \citet{b54} & \\
O$^{+}$ & \citet{b55}& \citet{b49}\\
             & \citet{b54}&\citet{b51}\\
O$^{++}$ & \citet{b54} & \citet{b42}\\
               & \citet{b50} &\\
S$^{+}$ & \citet{b47} & \citet{b52}\\
S$^{++}$ & \citet{b47} & \citet{b43}\\
Ar$^{++}$ & \citet{b45} & \citet{b43}\\
\hline
\hline
\end{tabular}
\end{center}
\end{table*}

\subsection{The nebular model}

\subsubsection{Abundances and plasma analysis}
The starting point to make a photoionization model for NGC 6888 is through the spectral analysis of the observations taken for different zones on this nebula.

As we said before, the fluxes obtained from the spectrum at position A contains not only the nebular emission but also the foreground emission. This foreground contribution is known from the spectra at positions B and C, where the flux values corresponding to the 15 km s$^{-1}$ component are similar. This enforces the assumption that these components have their origin in the interstellar medium surrounding the nebula and therefore, we can assume that these foreground fluxes are similar on the whole nebula NGC 6888.  By comparing the total fluxes from spectrum A with these values it is found that the foreground contribution is only around 1 per cent for all the identified lines, except for the faint [O {\small III}]$\lambda$$\lambda$ 4959,5007 lines where this flux contribution reaches 5 per cent. We can therefore assume that the integrated fluxes of the spectrum at this position represent well the nebular behaviour.

From the data in Table~\ref{tab:fluxes} we determined the electronic temperature $T_{e}$ and density $n_{e}$ from the available diagnostic ratios. In the case of positions B and C only the intense blueshifted components were used. In all the cases, only $T_{e}$ derived from the [N {\small II}] line flux ratio could be determined because the [O {\small III}] $\lambda$4363 line (crutial for determining $T_{e}$ [O {\small III}]) was not detected in the spectrum. The $n_{e}$ value comes from the usual [S {\small II}] diagnostic ratio because the lines coming from this ion are present in all the spectra. The temperature and density as well as the ionic abundances, derived by using these physical conditions and the line intensities from Table 2, are shown in Table~\ref{tab:par}. All these quantities as well as the errors were determined with the software {\small PyNeb 1.0.1b9} \citep{17} using the atomic data shown in Table~\ref{tab:atom}. The total abundances were only derived for position A where [O {\small II}] $\lambda$$\lambda$3726,3729 were detected. The ionization correction factors (ICF) by \citet{b14} were used with this purpose. The results are listed in Table~\ref{tab:par}.

In addition to our results, in Table~\ref{tab:par} we are including the results obtained by EV92 for their blueshifted component, and the data by FM12 corresponding to the inner zone observed by them, i.e., that labeled in their paper as B2. Both zones (EV92, and B2 FM12) are  clearly not associated with the rim region of NGC 6888 as it is shown in our Fig.~\ref{fig:slit} where their positions are marked. Also the results by M-D14, for their position in the rim, have been listed.

From these data we can make a direct comparison of the chemistry in different zones of NGC 6888, as obtained by different authors.

In Table~\ref{tab:par} it is found that our values of $T_{e}$[N {\small II}] electronic temperature and density for positions A, B, and C agree with the values derived by EV92, FM12 and M-D14. According to our observations, the plasma in position B$_{b}$ has the highest electronic temperature (although with a large uncertainty); similar values were reported by FM12 for the gas located farther than the edge of NGC 6888 and although their regions and our B position are not spatially close an explanation for the high temperature in position B could be that the main nebular emission comes from a low-density component of NGC 6888 if we assume this ring nebula is made of two gaseous components (see \S 4.2.2).

The abundances we obtain for position A in the rim are in good agreement with the previously published abundances (except for position B2 of FM12, see below). The fact that the values of O/H and N/H are compatible within the error bars for spatially different regions of the nebula indicates that the hypothesis of a chemically homogeneous nebulae need to be explored.

The abundances found for regions slightly farther than the edge of NGC 6888 (positions B1, E1, E2 and MB3 of FM12) are not taken into account because, according to FM12, there is no nitrogen enrichment there as this component was probably ejected during an early evolutionary stage of the star WR 136 as proposed by \citet{b40}.

The region B2 of FM12 presents an O/H abundance much larger than the values of the other positions. Note that FM12 did not obtain directly the electron temperature for the [O {\small III}] zone in their B2 region, but rather defined it using the relation (2) from \citet{b57} which was derived for other kind of objects as galaxies and H {\small II} regions, and given that NGC 6888 is an extented nebula which has a complex structure, probably made of two gaseous components, the changes in temperature we detect across the nebula cannot be assumed to follow the same behavior as found for global observations, since in addition we must deal with the ionization structure.

Therefore, their very low [O {\small III}] temperature value (not found by EV92 who directly determined it in a region close to B2) explains the high value they consequently obtain for O$^{++}$/H$^{+}$ ionic abundance and their high total O/H. If we use the T$_{e}$[O {\small III}] of EV92 the O$^{++}$ abundance is only 8.01 and the total abundance of oxygen is 8.29 which is in good agreement with the rest of the values shown in Table~\ref{tab:par}.

In addition FM12 obtain a low log~N/O ratio of $-$0.33 for this zone, which is also questionable, as this is contrary to the chemical enrichment predicted (log N/O $\sim$0.20) by the stellar evolution models (discussed below), where N-enrichment should be detected also near the star. This is actually the case for the also inner position observed by EV92 who obtained log N/O = 0.29. Note that the N/O abundance ratio is defined as equal to the N$^+$/O$^+$ ratio (ICF=1), while we will see in \S 5.3 and Fig.~\ref{fig:NO} that this is not the case for observations obtained through small slits in the inner zone of this extended nebula.
  
The values by FM12 could be indicating a chemically inhomogeneous nebula, as these authors suggested. However given the above discussion, it is possible instead to assume that the main structure of NGC 6888 has a homogeneous chemical abundance. The existence of such homogeneous nebulae are supported by the stellar evolution models for WR stars. For example, from Figure 16 of \citet{b24} we can see that along the evolution of a WR star, the atmospheric log N/O ratio reaches a high value of 0.20 in 10 000 years and remains almost constant for around 300 000 years, in the case of a non-rotational star with initial mass of 40 $M_{\odot}$.

In a similar way M-D14 followed the time evolution of the N/O ratio for NGC 6888, concluding that the material of the rim was ejected during the Red Super Giant phase highly nitrogen-enriched and the subsequent WR phase did not provide additional enrichment (see their Figure 5).

Therefore, given that the dynamical age of NGC 6888 has been estimated to be of about 20,000 - 40,000 yr \citep{b41}, the idea of a chemically homogenous nebula for NGC 6888 is consistent with the evolutionary models of a massive star becoming a WR star.

In the following we explore this possibility, by means of a photoionization model.

\begin{figure}
\includegraphics[width=0.6\textwidth]{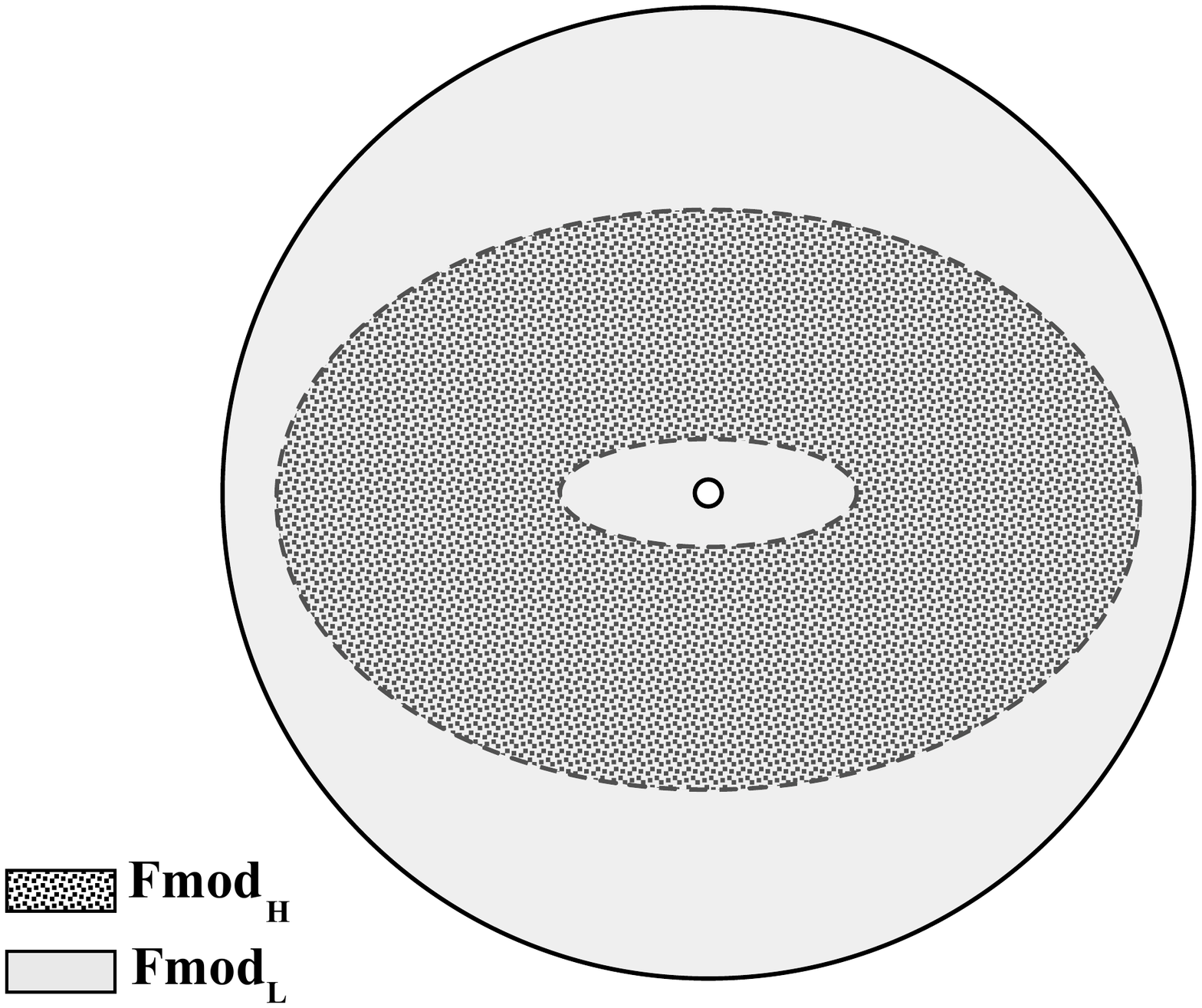}
 \caption{Scheme of the geometry assumed to build the photoinization model of NGC 6888. We can see the high-density and clumpy component ($Fmod_{H}$) as an ellipsoide covered by the spherical low-density component ($Fmod_{L}$). In the model, the edges of the ellipsoide match the size of the observed semi-axis of NGC 6888.}
 \label{fig:m2d}
\end{figure}

\begin{table}
\begin{center}
\caption{Abundance from the nebular and stellar models in 12+log(X/H)}
\label{tab:abun}
\begin{tabular}{crr}
\hline
\hline
Element & Nebular & \textit{star3}\\
\hline
He & 11.21&12.25\\
C & 8.86&7.84\\
N & 8.40&9.95\\
O & 8.20&8.76\\
Ne & 7.51&...\\
Si & ... &8.63\\
P & ... &8.01\\
S & 7.10&6.34\\
Ar & 6.41&...\\
Fe & 6.33&8.35\\
\hline
\hline
\end{tabular}
\end{center}
\end{table}

\subsubsection{Photoionization model of NGC 6888}

To reproduce the optical spectra of NGC 6888 we computed a photoionization model using the {\small pyCloudy} package (Morisset 2013) based on the 1D photoionization code Cloudy (Ferland 2008). It allows us to generate 2D images for each emission line from a pseudo-3D model and to apply a mask on them, reproducing the effect of the finite aperture through which the observations were performed. 

For our nebular model we used the version C10.0 of Cloudy, last described by \citet{b56}. We assumed, as it is apparent from the optical images \citep{b34,b19}, that the NGC 6888 projected morphology is the result of two gaseous components, as shown in Figure~\ref{fig:m2d}: one bright, granular region emitting mainly low ionization lines (e.g., [N {\small II}]), and the other one, more diffuse, mainly emitting the higher ionized emission lines (e.g., [O {\small III}]).

The denser (brighter) component has an ellipsoidal shape with an inner minor semi-axis of 80 arcmin, a minor to major semi-axis ratio of 0.33 with a H density of 400 cm$^{-3}$; the other component has a lower density of 1 cm$^{-3}$, and a spherical geometry with a central cavity with a radius of 20 arcmin. In addition we set a constant density profile for each component, with a filling factor $f_{mod_{L}}$ = 1.0 for the low-density component and $f_{mod_{H}}$ = 0.1 for the high-density component.

As we aim to build a consistent model of the system we used as ionizing sources the stellar atmosphere models of the star WR 136, as calculated with the {\small CMFGEN} code and described in the previous section. We therefore ensure to have well constrained stellar parameters, the ionization parameter and the ionization degree of the nebula through the stellar luminosity and temperature.

In a simple way, we calculate a chemically homogeneous model for NGC 6888 and we verify if observations can be reproduced.

The chemical abundances were first set to the values derived for position A in \S 4.2.1 (Table~\ref{tab:par}) which were completed with the abundances of C, Ne and Fe published by M-D14. Then the abundances were slightly modified until the photoionized model reproduces the observed line intensities.The final values which describe better the chemical composition in NGC 6888 are given in Table~\ref{tab:abun}.

The coordinates for the slit positions, from where the model integrated fluxes were obtained, were set in angular units with respect to the central star adopting a distance of 1.45 kpc for NGC 6888 \citep{b26}. In the optical images the star does not seem to be in the geometrical center of the nebula, but as we are only modelling the northwest part of it, the geometry of the model is an ellipsoid with the geometrical center on the central star position and whose edge matches the observed region.

The final intensities of the emission lines are obtained by summing up the contributions of each component, assuming that the high-density region weight on the total intensity is $w$. Thus: 

\begin{equation}
F_{mod} = f_{mod_{L}}F_{mod_{L}}(1-w) + f_{mod_{H}}F_{mod_{H}}w
\end{equation}

\noindent where $F_{mod_{L}}$ and $F_{mod_{H}}$ are the line intensity values from the low-density and high-density components, respectively. Once the abundances were fixed, only this weight value was changed, from one position to another, to fit the observed fluxes.

\begin{table*}
\caption{Comparison between the deredden observational and model fluxes (normalized to F$_{H_{\beta}}$ = 100) using \textit{star1} as ionizing source and \textit{w} = 1}
\label{tab:res1}
\begin{tabular}{llccccccccc}
\hline
\hline
$\lambda$ (\AA) & Ion & Fobs$_{A}$ & Fmod$_{A}$ & q$_{A}$& Fobs$_{B}$ & Fmod$_{B}$ & q$_{b}$& Fobs$_{C}$ & Fmod$_{C}$ & q$_{C}$\\
\hline
3726.03& [O II] & 52.6$\pm$14.3 & 52.5 & 0.00 &... &...&...&... &...&...\\
3728.81 &[O II] &62.6$\pm$9.5 & 61.7 & -0.13& ... &...&...&... &...&...\\
3970.07 &H$_{\epsilon}$ &16.8$\pm$6.1 & 15.1 & -0.35 & ... &...&...&... &...&...\\
4340.46& H$_{\gamma}$&47.1$\pm$5.7 & 46.3 & -0.13 & 48.8$\pm$13.4 & 46.8 & -0.17 & 47.7$\pm$3.5 & 47.0 & -0.17\\
4958.91 &[O III]&23.8$\pm$2.4 & 80.6 & 14.9 &97.0$\pm$6.2 & 121.5 & 2.91 & 52.1$\pm$4.5 & 136.0 & 12.51\\
5006.84 &[O III]&67.9$\pm$5.6 & 243.7 & 18.2 & 289.2$\pm$16.3 & 365.6 & 3.51 & 154.2$\pm$9.2 & 409.2 & 18.81\\
5015.67 &He I&4.7$\pm$1.8 & 2.8 & -1.74 & 6.0$\pm$1.8 & 4.2 & -1.62 & 4.7$\pm$3.1 & 4.2 & -0.21\\
5754.64 &[N II]&3.8$\pm$1.2 & 4.5 & 0.74 & 2.3$\pm$1.3 & 1.3& -1.33 & 2.3$\pm$0.9 & 1.1 & -1.22\\
5875.61 &He I&19.9$\pm$2.3 & 22.4 & 0.96 & 21.3$\pm$2.5 & 23.22 & 0.68 & 23.5$\pm$2.1 & 23.2 & -0.14\\
6312.10 &[S III]&...&...&...&2.2$\pm$1.3 & 4.4 & 1.67 &...&...&...\\
6548.04 &[N II]&168.4$\pm$11.5 & 125.8 & -4.25 & 37.1$\pm$2.8 & 32.1 & -2.83 & 70.9$\pm$3.5 & 24.3& -15.70\\
6562.80 &H$_{\alpha}$&285.9$\pm$18.3 & 291.4 & 0.30 & 286$\pm$14.1 & 290.1 & 0.16 & 286$\pm$10.8 & 289.2 & 2.00\\
6583.46 &[N II]&502.1$\pm$31.9 & 371.1 & -5.10 & 106.8$\pm$6.1 & 94.7 & -2.59 & 227.0$\pm$9.7 & 71.8 & -19.13\\
6678.15 &He I&6.4$\pm$1.2 & 6.3 & -0.13 & 6.4$\pm$2.8 & 6.6 & 0.10 & 5.2$\pm$0.7 & 6.5 & 2.79\\
6716.44 &[S II]&22.9$\pm$2.5 & 31.3 & 3.92 & 7.0$\pm$2.1 & 10.1 & 1.66 & 7.4$\pm$0.8 & 8.0 & 0.90\\
6730.81 &[S II]&19.9$\pm$1.9 & 28.4 & 4.73 & 5.7$\pm$2.4 & 9.8 & 1.73 & 6.3$\pm$1.2 & 7.8 & 2.36\\
7065.17 &He I& 3.2$\pm$1.0 & 3.9 & 0.84 & 6.0$\pm$2.6 & 4.4 & -0.88 & 6.4$\pm$0.8 & 7.83 & 0.66\\
7135.80 &[Ar III]&9.2$\pm$1.1 & 25.4 & 5.56 & 15.6$\pm$2.7 & 25.5 & 0.52 & 10.0$\pm$1.1 & 26.3 & 5.64\\ 
\hline
\hline
\end{tabular}
\end{table*}

\begin{table*}
\caption{Same as Tab. \ref{tab:res1}, but using \textit{star3} as ionizing source}
\label{tab:res2}
\begin{tabular}{llccccccccc}
\hline
\hline
 $\lambda$ (\AA)& Ion & Fobs$_{A}$ & Fmod$_{A}$ & q$_{A}$& Fobs$_{B}$ & Fmod$_{B}$ & q$_{b}$& Fobs$_{C}$ & Fmod$_{C}$ & q$_{C}$\\
\hline
3726.03& [O II] & 52.6$\pm$14.3 & 50.6 & -0.13 &... &...&...&... &...&...\\
3728.81 &[O II]&62.6$\pm$9.5 & 60.0 & -0.37& ... &...&...&... &...&...\\
3970.07 &H$_{\epsilon}$&16.8$\pm$6.1 & 15.1 & -0.36 & ... &...&...&... &...&...\\
4340.46 &H$_{\gamma}$&47.1$\pm$5.7 & 45.7 & -0.27 & 48.8$\pm$13.4 & 43.8 & -0.45 & 47.7$\pm$3.5 & 45.1 & -0.82\\
4958.91& [O III]&23.8$\pm$2.4 & 22.04 & -0.81 &97.0$\pm$6.2 & 68.0 & -3.03 & 52.1$\pm$4.5 & 51.5 & -0.13\\
5006.84 &[O III]&67.9$\pm$5.6 & 66.33 & -0.30 & 289.2$\pm$16.3 & 204.4 & -3.45 & 154.2$\pm$9.2 & 155.0 & 0.12\\
5015.67& He I&4.7$\pm$1.8 & 2.74 & -1.70 & 6.0$\pm$1.8 & 1.8 & -4.75 & 4.7$\pm$3.1 & 2.9 & -0.93\\
5754.64 &[N II]&3.8$\pm$1.2 & 3.63 & -0.12 & 2.30$\pm$1.3 & 1.0& -1.84 & 2.3$\pm$0.9 & 2.0 & -0.19\\
5875.61 &He I&19.9$\pm$2.3 & 22.7 & 1.21 & 21.3$\pm$2.5 & 19.8 & -0.62 & 23.5$\pm$2.1 & 21.8 & -0.97\\
6312.10 &[S III]&...&...&...&2.2$\pm$1.3 & 1.4 & -0.81 &...&...&...\\
6548.04 &[N II]&168.4$\pm$11.5 & 163.2 & -0.43 & 37.1$\pm$2.8 & 37.1 & 0.00 & 70.9$\pm$3.5 & 71.7& 0.14\\
6562.80 &H$_{\alpha}$&286$\pm$18.3 & 291.7 & 0.29 & 286$\pm$14.1 & 296.2 & 0.37 & 286$\pm$10.8 & 293 & 0.35\\
6583.46 &[N II]&502.1$\pm$31.9 & 481.7 & -0.71 & 106.7$\pm$6.1 & 109.6 & 0.70 & 227.0$\pm$9.7 & 211.48 & -1.10\\
6678.15 &He I&6.4$\pm$1.2 & 6.44 & 0.04 & 6.4$\pm$2.8 & 5.6 & -0.33 & 5.2$\pm$0.7 & 6.2 & 2.13\\
6716.44 &[S II]&22.9$\pm$2.5 & 15.87 & -3.51 & 7.0$\pm$2.1 & 2.1 & -4.58 & 7.4$\pm$0.8 & 3.9 & -6.75\\
6730.81 &[S II]&19.9$\pm$1.9 & 14.6 & -3.28 & 5.7$\pm$2.4 & 2.0 & -2.85 & 6.3$\pm$1.2 & 3.8 & -4.86\\
7065.17 &He I& 3.2$\pm$1.0 & 3.6 & 0.41 & 6.0$\pm$2.6 & 3.3 & -1.78 & 6.4$\pm$0.8 & 3.7 & -1.64\\
7135.80 &[Ar III]&9.2$\pm$1.1 & 10.5 & 1.14 & 15.6$\pm$2.7 & 8.79 & -0.15 & 10.0$\pm$1.1 & 10.2 & 0.24\\ 
\hline
\hline
\textit{w} (see text) & &7.6 $\times$ 10$^{-2}$& & & 1.7 $\times$ 10$^{-4}$ & & & 3.8 $\times$ 10$^{-4}$ &\\
\hline
\end{tabular}
\end{table*}

 The comparison between the model predictions and the observed dereddened line fluxes is presented in Table~\ref{tab:res1} and Table~\ref{tab:res2} corresponding to the photionization model results by using \textit{star1} and \textit{star3} SED's as ionizing source, respectively. The \textit{q} parameter, in these tables, shows the goodness of our model and is defined following \citet{b30} as:\\

\begin{equation}
q =\frac{log(F_{mod})-log(F_{obs})}{t}
\end{equation}

\noindent where $F_{mod}$ is the value returned by the model, $F_{obs}$ is the observed flux and $t$ is the accepted tolerance in dex of this observable. We define the tolerance to be\\

\begin{equation}
t =log\left(1+\frac{\delta F_{obs}}{F_{obs}}\right)
\end{equation}

\noindent where $\delta F_{obs}$ is the measured error. When a line flux is well reproduced the \textit{q} value is between $-$1.0 and 1.0.

The results in Table~\ref{tab:res1} show that, under the assumptions made for this model and using \textit{star1} as ionizing source, the observed line fluxes are not well reproduced. In particular the lines from highly ionized elements (O$^{++}$, Ar$^{++}$) are extremely overpredicted. The best case is for the values in position A where only half of the whole line fluxes (the low excitation ones) detected in this region are well reproduced. The model results are worse for the other positions. 

The fact that the ionization of the nebula is strongly overpredicted could be due to: 1) a too high ionization parameter $U$\footnote{$U = \frac{Q_{\ast}}{4\pi n_{H}R^{2}}$ where $Q_{\ast}$ is the ionizing photon rate, $n_{H}$ the hydrogen number density and $R$ the distance to the ionizing source.}, 2) a too high stellar effective temperature. Notice that this model was obtained using only the high density component ($w=1$). Adding any contribution of the low density component would made the situation even worse.
  
As $R$, $Q_{\ast}$ and $n_H$ are well constrained through the \textit{HIPPARCOS} distance to the object, its observed angular size, the observed stellar luminosity and the nebular density coming from the plasma diagnostic, then it follows that $U$ is also well constrained. The only remaining way to reproduce the observed low value of the [O {\small III}]/[O {\small II}] line ratio is to consider a star with a temperature lower than the value derived for model \textit{star1}, i.e., we must not deal with the number of ionization photons but rather with their energy. 

The fit improves very much when we use the model \textit{star3} which has $T_{\ast} =$ 70 000 K. The ionization of the dense gas is even lower than observed, leading to the full use of our hypothesis of the 2-components model. The predicted emission lines from the corresponding photoionization model are compared with observations in Table~\ref{tab:res2}. It can be seen that almost all the fluxes are well reproduced, indicating that the star WR 136 emits the UV ionizing photons corresponding to a star with this low effective temperature.

\section{Discussion}

\subsection{Effective temperature and mass-loss rate of WR 136}

As it can be seen in Figure~\ref{fig:stelspec} and it was discussed in \S 4.1, at least three stellar models, (\textit{star1, star2} and \textit{star3}) with different $T_{\ast}$ and therefore a different $R_{t}$ fit similarly well the stellar observations, in particular the stellar continuum and several emission lines. The differences among these models appear mainly in the He lines, where the He {\small II}/He {\small I} line ratio is well reproduced when considering an effective temperature of 110 000 K but it is understimated when the effective temperature is reduced, i.e., at a lower stellar temperature the stellar He {\small I} lines are stronger and the stellar He {\small II} lines become weaker than observed. As we discussed before, it seems that there is not a unique solution for the stellar model.

\begin{figure}
\includegraphics[width=0.5\textwidth]{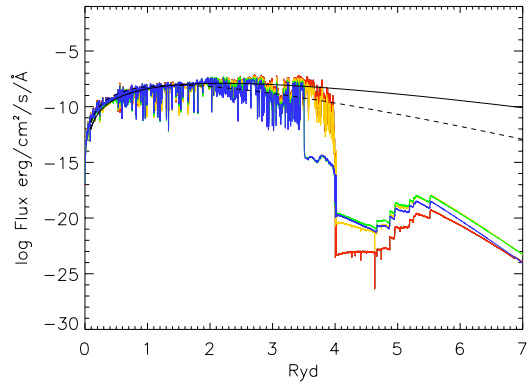}
 \caption{Comparison of the ionizing region of the spectra corresponding to stellar models with diferent effective temperature. The flux is not dereddened and is shown as emitted at a distance of a 1 kpc. Colors are as in Figure~\ref{fig:steli}. The black solid line and the dashed line correspond to blackbodies of 70 000 and 46 000 K, respectively, according to the temperatures $T_{\ast}$ and $T_{2/3}$ of model \textit{star3}.}
\label{fig:spectrauv}
\end{figure}

In Figure~\ref{fig:spectrauv} we show a comparison between the UV region of the stellar models from Table~\ref{tab:w136}. We also overplot the UV behaviour of two blackbodies with temperatures of 70 000 and 46 000 K which correspond to the temperatures $T_{\ast}$ and $T_{2/3}$ of model \textit{star3}. 

We can see that the key difference between the stellar models comes from the far UV region of the spectra, whose UV photons are absorbed by the nebula (between 1.0 and $\sim$ 7.0 Ryd). This difference is well appreciated when we compare the models with the blackbodies. From this figure it is clear that a blackbody with the same temperature than the modelled spectra, 70 000 or even 46 000 K, shows an excess of UV photons which would lead to a much more ionized nebula. It is also evident the high absorption of the UV photons emitted by the star, due to the presence of the stellar wind. 

Owing to there are not stellar observations in the far UV to discriminate which stellar model is reproducing the real spectral energy distribution of the star WR 136, it is necessary to add an external restriction. In this case, such additional restriction is the associated nebula NGC 6888 that is highly sensitive to this UV ionizing radiation.

In addition, the combination of the temperature and mass-loss effects in the stellar wind, considered for this model, can reproduce the line profiles coming from recombination process. Therefore, if we are close to a good value for $T_{\ast}$, it is the same for the mass-loss rate, $\dot{M}$. This is important because $\dot{M}$ represents a good constrain for the hydrodynamical simulations. 

As in this study we have found a well constrained value for $\dot{M}$ (log $\dot{M}$ = $-$4.95), which is similar to the values found by other authors (Table~\ref{tab:w136}), we can rule out the assumption made recently by \citet{b28} where they conclude that it is necessary a $\dot{M}$ 3 - 4 times lower than observed for this type of stars to reproduce the X-ray flux of the emitting hot bubble.

\begin{figure*}
\includegraphics[width=1\textwidth]{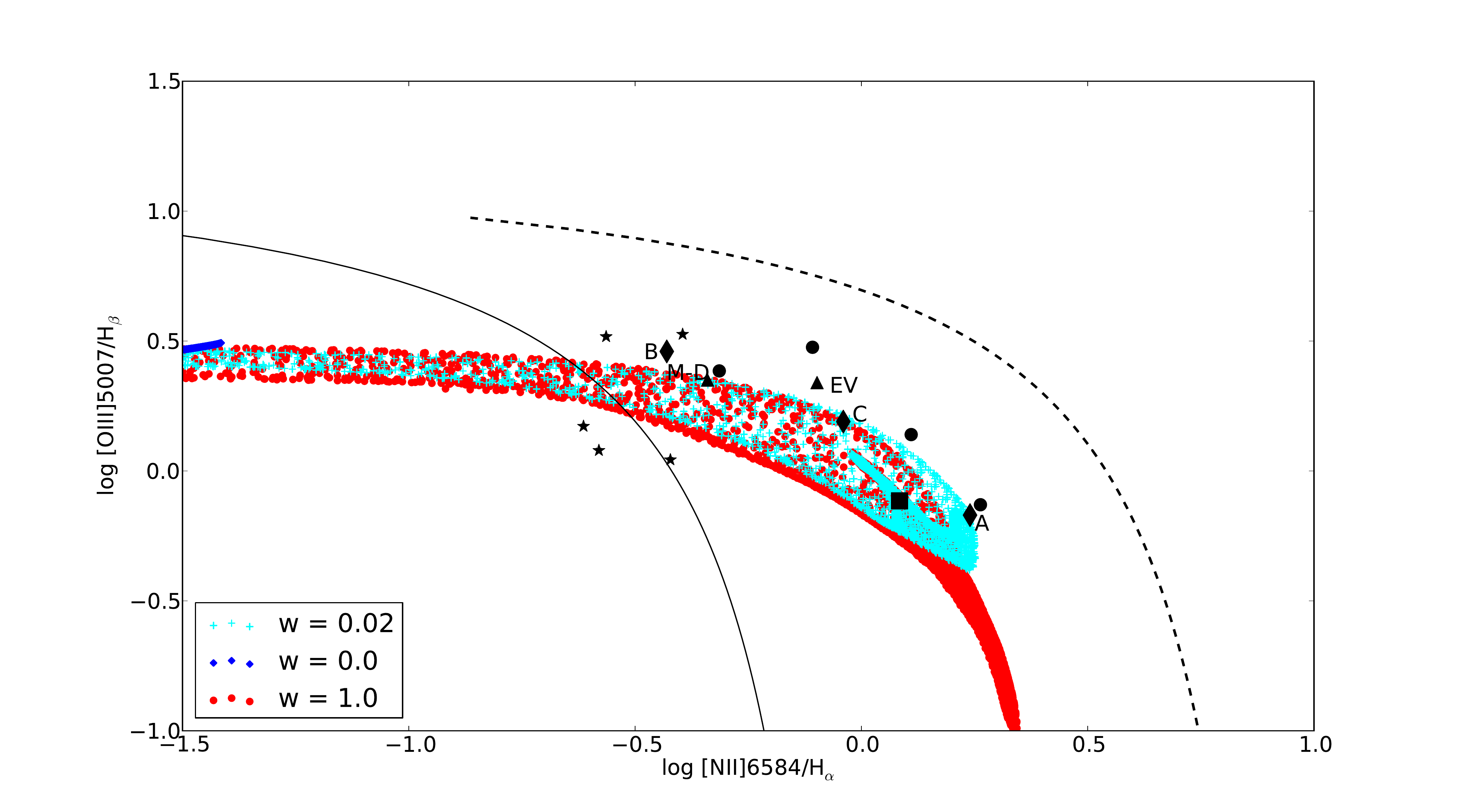}
 \caption{Plasma diagnostic diagram [O {\small III}]/H$\beta$ vs. [N {\small II}]/H$\alpha$. The solid curve corresponds to the diagram in \citet{b32} and the dashed curve is the same one but shifted by 0.96 dex given the difference between the N/O values in that work and in ours. For three different \textit{w} values, all the points that make the photoionization model of NGC 6888, using \textit{star3} as ionization source, are shown. The observed ratios are marked as black diamonds and the black square represents the values of integrated fluxes coming from the whole nebula. The observations by FM12 are also shown: stars represent the N-poor regions in the immediate outer zone of NGC 6888 while the black points are observations of the main structure of the nebula with N-rich abundance. Observations by EV92 and M-D14 are also show in the diagram as black triangles.} 
\label{fig:BPT}
\end{figure*}

\subsection{Is NGC 6888 mainly photoionized?}

The presence of a very fast stellar wind coming from the star WR 136 could imply that part of the ionization of the gas is due to shocks. Diagrams made by \citet{b27} and \citet{b31} are usually employed to determine if a line emitting region is mainly ionized by OB stars (like H {\small II} regions) or by other source (AGN, LINERS, shocks). See for example \citet{b29} for the case of shocks. 

In Figure \ref{fig:BPT} we show an [O\,{\small III}]\,5007/H$\beta$ vs. [N\,{\small II}] 6584/H$\alpha$ diagram. The solid line shows the separation between global photoionization models for Star Forming regions (NSF galaxies) and AGNs as defined by \citet{b32}. The dashed line represents the same separation, but shifted on the log ([N {\small II}] 6584/H$\alpha$) axis by 0.96 dex, which is the difference between the N/O adopted by \citet{b32} and the N/O of NGC 6888 deduced in this work. The three black diamonds correspond to observations for positions A, B and C and are obviously located to the right side of the solid line, but well inside the photoionized zone with N/O ratio similar to NGC 6888 value.

The effect of a limited aperture size, which is not including the whole nebula is illustrated by the position of the color points in our Fig.~\ref{fig:BPT}. They show the emission line ratios of every spaxel of our 2D images obtained from our models. The colors correspond to different values of $w$ (the weight of the dense component), the blue symbols are obtained for the case of considering only the low-density component, and the red ones for the case of considering only the high-density component, the green symbols being an intermediate situation. As we can see, the points spread in an ample region of the diagram. The black square symbol is the position in the diagram of the global model, as if it were fully included in the slit.

Therefore if just a small portion of a whole nebula is observed, one must be careful in using diagnostic diagrams to decide if the emission of a nebula is caused by the presence of shocks, because as we demonstrated in Fig.~\ref{fig:BPT} only photoionization and an enhanced N abundance are sufficient to obtain points located out of the regions represented by global photoionized models (solid line).

\subsection{Oxygen and nitrogen abundances in NGC 6888}

To deeper investigate our assumption of chemical homogeneity of NGC 6888, in Fig.~\ref{fig:BPT} we added the observations by EV92, M-D14. It is found that they lie inside or close to our model points. Also we added the observations by FM12 (marked as black dots and stars) resulting from their 1D analysis of different regions of NGC 6888. We can see that their points show a large spread in this diagram. To explain the behaviour of their observations and to derive values of the N and O abundances, FM12 performed a grid of photoionization models using the code Cloudy varying the ionization parameter and the N/O ratio. As they used global models for each observation without taking into account that they were observing only a part of the nebula they concluded that there are important differences in the N/O ratio across NGC 6888. 

However, our figure shows that this is not necessarily the case. In particular, let us focus on the data of FM12 marked as black dots, which correspond to their regions inside NGC 6888 with large nitrogen enrichment. We can see that all these points, but the point labeled X2 in their paper, are close to our model points and are following their trend. The rest of the points (stars) show a large dispersion compared to our model points and it is not expected for the model to reproduce their positions because they belong to regions located out from the edge of NGC 6888 with low (solar) nitrogen abundance.

Therefore, Figure~\ref{fig:BPT} illustrates the fact that all the observed points {\it within} the nebula can be reproduced by a rather simple chemically homogeneous model.

\begin{figure}
\includegraphics[width=1.0\linewidth]{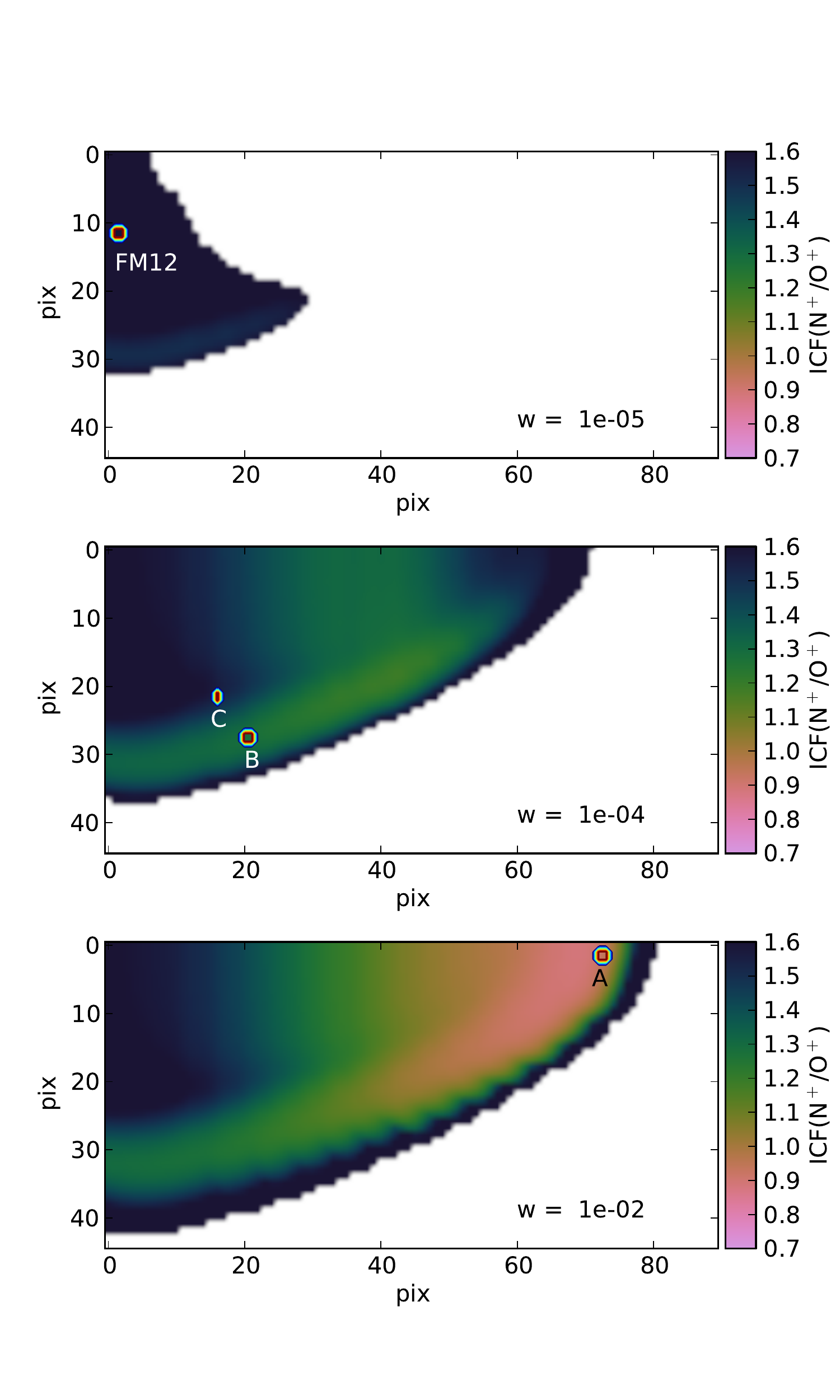}
 \caption{Map of the ICF for the N$^{+}$/O$^{+}$ ratio from a quadrant of our best photoionization model which is chemically homogeneous. The star is assumed to be at the origin of the map. Three different values of \textit{w} were used to show how the values of the ICF changes over the whole structure of nebula. Our positions and the position of FM12 are shown.}
\label{fig:NO}
\end{figure}

To continue the discussion about the chemical homogeneity in NGC 6888, we analyse the behaviour of the ICF used to determine the N/O ratio. Let us remember that it is generally assumed that N$^{+}$/O$^{+}$ is equal to N/O.

From our photoionization model, we are able to determine in each region of the nebula the ICF needed for the N$^{+}$/O$^{+}$ ratio to obtain the N/O value. We show in Figure~\ref{fig:NO} a map of this ICF for three different values of \textit{w}: 10$^{-5}$, 10$^{-4}$ and 10$^{-2}$. It is evident that for a given \textit{w}, this ICF varies over the nebula showing a difference as large as a factor of two. In adittion, we can see that the value of this ICF changes as the contribution of the high-density component is modified. 

For example, in the case when \textit{w} = 10$^{-2}$, the ICF for position A is around 1.0. In general, we can see in Figure~\ref{fig:NO} that for positions closer to the star the ICF is around 1.5 or larger. Therefore, in the zones near to the star an ICF greater than one is needed to estimate the N/O ratio. Hence, this indicates that N/O could be understimated when it is simply assumed that N$^{+}$/O$^{+}$ is equal to N/O. In this way, to calculate the nitrogen abundance for position B2 of FM12 and for EV92 a better value for this ICF is 1.5. Once this ICF is used for these positions the N/O ratio increases relative to the values in Table~\ref{tab:par}, to 0.47 for EV92 and $-$0.15 for B2 FM12. Therefore these values are still discrepant. In particular, for the position B2 the N/O ratio reaches a value of $-$0.15 lower than the others for the rim, which is contrary to the predictions of stellar evolution models.

The differences seen of the ICF in different zones of the map are due to the combination of two effects: the ionization structure of the nebula and the relative contribution to the fluxes of the high density clumpy region.

A similar problem with the ICFs is affecting the results of \citet{b38} who used \textit{Herschel} observations of the [O {\small III}] 88 $\mu$m and [N {\small III}] 57 $\mu$m lines, emitted in two positions of NGC 6888, to determine the N/O abundance ratio under the assumption that N$^{++}$/O$^{++}$ is equal to N/O. For NGC 6888 this is a very rough approximation because only a small fraction of the nebula was included in their slits and \cite{b38} did not take into account the nebular ionization structure, which indicates that the ion N$^{++}$ is only dominant over N$^{+}$ in the zones near to the star.

The previous discussion illustrates very well that detailed modelling (even the simplest as we do here: 2 almost roundish homogeneous components) is a very powerful way to identify effects that, if not considered, can easily drive the observer to wrong conclusions.

\section{Conclusions}

We computed several atmosphere models using the code {\small CMFGEN} to reproduce the stellar observations of WR 136. The characteristics of the star such as $L$, $\dot{M}$ and $V_{\infty}$ agree with the values found by other authors, i.e., HGL06 and \citet{b3}. We also derived the stellar abundances of He, C , N, O, Si, P, S and Fe (listed in Table~\ref{tab:abun}) which agree with HGL06's model. 

We find that models with very different effective temperature, $T_{\ast}$ from 70 000 K to 110 000 K, can reproduce fairly well the stellar continuum and several emission lines, although the He lines, indicative of temperature, are worse reproduced by the cooler models. Therefore there is a degeneracy in atmosphere models, than apparently, as pointed out by \citet{b9}, is due to our misunderstanding of the structure of the wind velocity field.

To decide which stellar model produces the correct UV amount of ionizing photons coming from WR 136, we computed a photoionization model for the associated ring nebula NGC 6888, using the SED produced by the stellar models. This led us to conclude that WR 136 follows better the spectral behaviour of a star with a $T_{\ast}$ of 70 kK.

The photoionization model was built assuming chemical homogeneity and including two components: a low-density region enclosing a denser region.
 
When these two gaseous components are considered and assuming only photoionization, we were capable to reproduce the observed ionization structure in both shells, contrary to the presence of shocks as the probable mechanism producing the observed ionization structure of NGC 6888. We cannot immediatly rule out the importance of shock effects, but given that some diagnostic diagram such as the [O {\small III}] 5007/H$\beta$ vs. [N {\small II}] 6584/H$\alpha$ indicate emission of H {\small II} type, and that we can reproduce several emission features of the nebula by only assuming photoionization, we established that the photoionization heating is more important than the presence of shocks. This latter idea was already mentionned by EV92 combining the diagnostic diagrams of \citet{b22} with their observations in a zone close to the central star and shown in Figure~\ref{fig:slit}.
 
The nebular abundances derived from our photoionization model are similar to the values derived from our position A and the values reported by EV92 and M-D14. 

It is worth to notice that both the stellar atmosphere of WR 136 and the nebular plasma present a large N- and C-enhancement and an O depletion, as it can be seen in Table~\ref{tab:abun}. This contributes to the idea that the observed gas was ejected by the central star through the mass-loss driven by the stellar wind.

Considering only photoionization and a chemically homogeneous nebula, we were able to explain the pretended differences found in the N/O ratio (derived from N$^{+}$/O$^{+}$ or N$^{++}$/O$^{++}$ in different zones of the nebula) by FM12 and \citet{b38}, who proposed a chemically inhomogeneous nebula. These pretended differences are due to effects of the ionization structure within the nebula. With our simple model we show that the ICFs for global nebulae should not be use for observations of small zones observed in extended nebulae.

From our results we have learned that there are several stellar models with different set of values that could reproduce the observables, like the models \textit{star1} and \textit{star3} presented in this study. Therefore external constraints are crucial to determine which atmosphere models represent the real stellar properties. The nebulae around stars can help to deal with the degeneracy mentioned above.

Finally, in this work we show the advantages of modelling simultaneously the stellar and nebular spectra of a Wolf-Rayet star plus nebula system. We verify whether a stellar atmosphere model of the star WR 136 provides a good description of the stellar observations and is able to satisfactorily reproduce the nebular characteristics such as the ionization degree. This allow us to better constraint all the physical parameters of both the star and the nebula, including their physical conditions and their chemical abundances.

\section{Acknowledgements}
This work received partial finantial support from UNAM PAPIIT IN109614 and CONACyT CB-2010/153985.
The FUSE and IUE data presented in this paper were obtained from the Multimission Archive at the Space Telescope Science Institute (MAST). STScI is operated by the Association of Universities for Research in Astronomy, Inc., under NASA contract NAS5-26555. Support for MAST for non-HST data is provided by the NASA Office of Space Science via grant NAG5-7584 and by other grants and contracts. We thank G. Koenigsberger, A. Peimbert and J. Garc\'ia-Rojas for their helpful comments.

\label{lastpage}


\begin{thebibliography}{99}
\bibitem[\protect\citeauthoryear{Aggarwal \& Keenan}{1999}]{b42} Aggarwal K. M., Keenan F. P. 1999, ApJS, 123, 311
\bibitem[\protect\citeauthoryear{Allen et al.}{2008}]{b29} Allen Mark G., Groves Brent A., Dopita Michael A., Sutherland Ralph S., Kewley Lisa J., 2008, ApJS, 178, 20
\bibitem[\protect\citeauthoryear{Baldwin, Phillips \& Terlevich}{1981}]{b31} Baldwin J. A., Phillips M. M., Terlevich R., 1981, PASP, 93, 5	
\bibitem[\protect\citeauthoryear{Cardelli, Clayton \& Mathis}{1989}]{b1} Cardelli Jason A., Clayton Geoffrey C., Mathis John S., 1989, ApJ, 345,245
\bibitem[\protect\citeauthoryear{Chu, Treffers, \& Kwitter}{1983}]{b33} Chu Y.-H., Treffers R. R., Kwitter K. B. 1983, ApJS, 53, 937
\bibitem[\protect\citeauthoryear{Crowther}{2007}]{b2} Crowther P. A. 2007, ARA\&A, 45, 177
\bibitem[\protect\citeauthoryear{Crowther \& Smith}{1996}]{b3} Crowther P. A., Smith L. J. 1996, A\&A, 305, 541
\bibitem[\protect\citeauthoryear{Eenens \& Williams}{1994}]{b4} Eenens P.R.J., Williams P. M., 1994, MNRAS, 269, 108
\bibitem[\protect\citeauthoryear{Esteban \& V\'ilchez}{1992, hereafter EV92}]{b5} Esteban C.,  V\'ilchez J. M. 1992, ApJ, 390, 536, (EV92)
\bibitem[\protect\citeauthoryear{Ferland et al.}{1998}]{b6} Ferland G. J., Korista K. T., Verner D. A., Ferguson J. W., Kingdon J. B.,  Verner E. M., 1998, PASP, 110, 761
\bibitem[\protect\citeauthoryear{Ferland et al.}{2013}]{b56} Ferland G. J., Porter R. L., van Hoof P. A. M., Williams R. J. R., Abel N. P., Lykins M. L., Shaw G., Henney W. J., Stancil P. C. 2013, RMAA, 49, 13
\bibitem[\protect\citeauthoryear{Fern\'andez-Mart\'in et al.}{2012, hereafter FM12}]{b35} Fern\'andez-Mart\'in A., Mart\'in-Gordon D., V\'ilchez J. M., P\'erez Montero E., Riera A., S\'anchez S. F., 2012, A\&A, 541, 119, (FM12)
\bibitem[\protect\citeauthoryear{Galavis, Mendoza \& Zeippen}{1995}]{b43} Galavis M.E., Mendoza C., Zeippen C. J. 1995, A\&AS 111, 347
\bibitem[\protect\citeauthoryear{Galavis, Mendoza \& Zeippen}{1997}]{b44} Galavis M. E., Mendoza C., Zeippen C. J. 1997, A\&AS, 123, 159
\bibitem[\protect\citeauthoryear{Garc\'ia-Segura, Langer \& Mac Low}{1996}]{b40} Garc\'ia-Segura G., Langer N., Mac Low M.-M. 1996, A\&A, 316, 133
\bibitem[\protect\citeauthoryear{Gray}{1999}]{b7} Gray R. O. 1999, ascl.soft10002G
\bibitem[\protect\citeauthoryear{Gruendl et al.}{2000}]{b34} Gruendl R. A., Chu Y.-H., Dunne B. C., Points S. D. 2000, AJ, 120, 2670
\bibitem[\protect\citeauthoryear{Hamann, Grafener \& Liermann}{2006, hereafter HGL06}]{b8} Hamann W.-R., Grafener G., Liermann A., 2006, A\&A, 457, 1015, (HGL06)
\bibitem[\protect\citeauthoryear{Hillier}{1991}]{b9} Hillier D. J., 1991, IAUS, 143, 59
\bibitem[\protect\citeauthoryear{Hillier \& Miller}{1998}]{b10} Hillier D. J., Miller D. L. 1998, ApJ, 496, 407
\bibitem[\protect\citeauthoryear{Hillier et al.}{2003}]{b11} Hillier D. J., Lanz T., Heap S. R., Hubeny I., Smith L. J., Evans C. J., Lennon D. J., Bouret J. C. 2003, ApJ, 588, 1039
\bibitem[\protect\citeauthoryear{Ignace et al.}{2001}]{b12} Ignace R., Cassinelli J. P., Quigley M., Babler B. 2001, ApJ, 558, 771
\bibitem[\protect\citeauthoryear{Ignace, Quigley \& Cassinelli}{2003}]{b13} Ignace R., Quigley M. F., Cassinelli J. P., 2003, ApJ, 596, 538
\bibitem[\protect\citeauthoryear{Kauffmann et al.}{2003}]{b32} Kauffmann G., Heckman T. M., Tremonti C., Brinchmann J., Charlot S., White S. D. M., Ridgway S. E., Brinkmann J., Fukugita M., Hall P. B., Zeljko I., Gordon T. R., Donald P. S. 2003, MNRAS, 346, 1055
\bibitem[\protect\citeauthoryear{Kaufman \& Sugar}{1986}]{b45}Kaufman V., Sugar J. 1986, JPCRD, 15, 321
\bibitem[\protect\citeauthoryear{Kingsburgh \& Barlow}{1994}]{b14} Kingsburgh R. L., Barlow M. J. 1994, MNRAS, 271, 25
\bibitem[\protect\citeauthoryear{Kwitter}{1981}]{b37} Kwitter K. B., 1981, ApJ, 245, 154
\bibitem[\protect\citeauthoryear{Lamers \& Cassinelli}{1999}]{b15} Lamers H. J. G. L. M., Cassinelli J. P. Introduction to stellar winds / Cambridge U Press, 1999
\bibitem[\protect\citeauthoryear{Levine \& Chakrabarty}{1994}]{b16} Levine S., Chakrabarty D. 1994, Tevhnical Report MU-94-04, Instituto de Astronom�a, Universidad Nacional Aut\'onoma de M\'exico.
\bibitem[\protect\citeauthoryear{Luridiana, Morisset \& Shaw}{2015}]{17} Luridiana V., Morisset C., Shaw R. A., 2015, A\&A, 573,42
\bibitem[\protect\citeauthoryear{Mendoza}{1983}]{b46} Mendoza C. 1983, I.A.U. Symp. No. 103, p. 143
\bibitem[\protect\citeauthoryear{Mesa-Delgado et al.}{2014, hereafter M-D14}]{b18} Mesa-Delgado A., Esteban, C., Garc\'ia-Rojas J. Reyes-P\'erez J. Morisset C., Bresolin F., 2014, ApJ, 785, 100, (M-D14)
\bibitem[\protect\citeauthoryear{Moore, Hester \& Scowen}{2000}]{b19} Moore Brian D., Hester J. Jeff, Scowen Paul A., 2000, AJ, 119, 2991
\bibitem[\protect\citeauthoryear{Morisset}{2013}]{b20} Morisset C. 2013. sites.google.com/site/cloudy3d
\bibitem[\protect\citeauthoryear{Morisset \& Georgiev}{2009}]{b30} Morisset C., Georgiev L., 2009, A\&A, 507, 1517
\bibitem[\protect\citeauthoryear{P\'erez-Montero \& Contini}{2009}]{b57} P\'erez-Montero E., Contini T. 2009, MNRAS, 398, 949
\bibitem[\protect\citeauthoryear{Podobedova, Kelleher \& Wise}{2009}]{b47} Podobedova L. I., Kelleher D. E., Wiese W. L. 2009, JPCRD, 38,171
\bibitem[\protect\citeauthoryear{Porter et al.}{2013}]{b48} Porter R. L., Ferland G. J., Storey P. J., Detisch M. J. 2013, MNRAS 433, 89
\bibitem[\protect\citeauthoryear{Pradhan et al.}{2006}]{b49} Pradhan A. K., Montenegro M., Nahar S. N., Eissner W. 2006, MNRAS, 366, 6
\bibitem[\protect\citeauthoryear{Prinja, Barlow \& Howarth}{1990}]{b21} Prinja Raman K., Barlow M. J., Howarth Ian D. 1990, ApJ, 361,607
\bibitem[\protect\citeauthoryear{Sabbadin, Minello \& Bianchini}{1977}]{b22} Sabbadin F., Minello S., Bianchini A. 1977, A\&A, 60, 147
\bibitem[\protect\citeauthoryear{Schmutz, Hamann \& Wessolowski}{1989}]{b23} Schmutz W., Hamann W.-R., Wessolowski U., 1989, A\&A, 210,236
\bibitem[\protect\citeauthoryear{Skinner et al.}{2010}]{b36} Skinner Stephen L., Zhekov Svetozar A., G\"{u}del Manuel, Schmutz Werner, Sokal Kimberly R., 2010, AJ, 139, 825
\bibitem[\protect\citeauthoryear{Stock \& Barlow}{2014}]{b38} Stock D. J., Barlow M. J., 2014, MNRAS, 441, 3065
\bibitem[\protect\citeauthoryear{Storey \& Zeippen}{2000}]{b50} Storey P.J., Zeippen C.J., MNRAS, 312, 813, 2000
\bibitem[\protect\citeauthoryear{Tayal}{2007}]{b51} Tayal S. S., 2007, ApJS, 171, 331
\bibitem[\protect\citeauthoryear{Tayal}{2011}]{b53} Tayal S. S. 2011, ApJS, 195, 12
\bibitem[\protect\citeauthoryear{Tayal \& Zatsarinny}{2010}]{b52} Tayal S. S., Zatsarinny, O. 2010, ApJS, 188, 32
\bibitem[\protect\citeauthoryear{Toal\'a \& Arthur}{2011}]{b24} Toal\'a J. A., Arthur S. J. 2011. ApJ, 737, 100
\bibitem[\protect\citeauthoryear{van der Hucht}{2001}]{b39} van der Hucht K.A., 2001, New Astron. Rev., 45, 135
\bibitem[\protect\citeauthoryear{van der Hucht, Cassinelli \& Williams}{1986}]{b25} van der Hucht K.A., Cassinelli J.P., Williams P.M., 1986, A\&A, 168, 111
\bibitem[\protect\citeauthoryear{van Leeuwen}{2007}]{b26} van Leeuwen F., 2007, A\&A, 474, 653.
\bibitem[\protect\citeauthoryear{van Marle, Langer \& Garc\'ia-Segura}{2005}]{b41} van Marle A. J., Langer N., \& Garc\'ia-Segura G. 2005, A\&A, 444, 837
\bibitem[\protect\citeauthoryear{Veilleux \& Osterbrock}{1987}]{b27} Veilleux S., Osterbrock D., E., 1987, ApJS, 63, 295
\bibitem[\protect\citeauthoryear{Wiese, Fuhr \& Deters}{1996}]{b54} Wiese W. L., Fuhr J. R., Deters T. M., JPCRD Monograph(7) pp. 1-522 (1996)
\bibitem[\protect\citeauthoryear{Zeippen}{1982}]{b55} Zeippen C. J. 1982, MNRAS, 198, 111
\bibitem[\protect\citeauthoryear{Zhekov \& Park}{2011}]{b28} Zhekov S. A., Park S. 2011, ApJ, 728, 135
\end{thebibliography}
\end{document}